%% file: dsfe-301.tex
\PassOptionsToPackage{table}{xcolor}
\documentclass{dsfe}

\usepackage{txfonts}
\def\typeofarticle{Research article}
\def\currentvolume{5}
\def\currentissue{3}
\def\currentyear{2025}

\def\ppages{387--418.}
\def\DOI{10.3934/DSFE.2025016}
\def\Received{25 November 2024}
\def\Revised{16 June 2025}
\def\Accepted{23 July 2025}
\def\Published{29 August 2025}

\usepackage{graphicx} 
\usepackage{subfigure}
\usepackage[ruled]{algorithm2e}
\usepackage{amsmath}
\usepackage[makeroom]{cancel}
\usepackage[normalem]{ulem}
\usepackage{longtable}
\usepackage[table]{xcolor}
\usepackage{multirow}
\usepackage{booktabs}
\usepackage{float}
\usepackage{bigfoot}
\DeclareMathOperator*{\argmin}{arg\,min}
\usepackage{tikz}
\usepackage{bm}
\usepackage{relsize}
\usepackage{microtype}
\usepackage[natbibapa]{apacite}
\usepackage{doi}

\emergencystretch=\maxdimen
\theoremstyle{plain}

\theoremstyle{definition}

\theoremstyle{remark}

\begin{document}

\title{Automated regime classification in multidimensional time series data using sliced Wasserstein k-means clustering}

\author{%
  Qinmeng Luan\affil{1}
  and
  James Hamp\affil{1,2,}\corrauth
}

\shortauthors{the Author(s)}

\address{%
  \addr{\affilnum{1}}{Citigroup, London, UK}
  \addr{\affilnum{2}}{Data Science Institute, London School of Economics, London, UK}}

\corraddr{Email: james.o.hamp@gmail.com; Tel: +447896844693.
}

\begin{abstract}
Recent work has proposed Wasserstein k-means (Wk-means) clustering as a powerful method to classify regimes in time series data, and one-dimensional asset returns in particular.
In this paper, we begin by studying in detail the behaviour of the Wasserstein k-means clustering algorithm applied to synthetic one-dimensional time series data.
We extend the previous work by studying, in detail, the dynamics of the clustering algorithm and how varying the hyperparameters impacts the performance over different random initialisations. 
We compute simple metrics that we find to be useful in identifying high-quality clusterings.
We then extend the technique of Wasserstein k-means clustering to multidimensional time series data by approximating the multidimensional Wasserstein distance as a sliced Wasserstein distance, resulting in a method we call `sliced Wasserstein k-means (sWk-means) clustering'.
We apply the sWk-means clustering method to the problem of automated regime classification in multidimensional time series data, using synthetic data to demonstrate the validity and effectiveness of the approach. 
Finally, we show that the sWk-means method is able to identify distinct market regimes in real multidimensional financial time series, using publicly available foreign exchange spot rate data as a case study.
We conclude with remarks about some limitations of our approach and potential complementary or alternative approaches.
\end{abstract}

\keywords{ \textls[-20]{time series; regime classification; market regimes; Wasserstein metric; unsupervised learning}\\
\vspace{0.4cm}
}

\maketitle

\section{Introduction}

The analysis of time series is of central importance in many domains, not least in finance, where asset prices and other economic time series are studied in order to quantify past and current macroeconomic conditions and/or identify investment opportunities, for example.
Generally, it can be useful to characterise the behaviour of time series in terms of `regimes', which are periods during which the statistical properties of the time series remain similar, compared with other periods. 
In finance, such regimes are called `market regimes', which might correspond to periods of bullish/bearish performance in equities, periods of high/low inflation, or periods of high/low volatility in foreign exchange (FX) rates, to name but a few examples. 
In general, regimes are also characterised by the joint behaviour of multiple time series. 
Most simply, the joint behaviour can be characterised in terms of correlations, with regimes corresponding to periods of different correlations between the time series, in addition to their marginal behaviour. In finance, one might be interested in studying the joint distributions of time series either within a given asset class or between different asset classes.

A key objective is the ability to rapidly and automatically identify regimes in time series, including multidimensional time series. 
Recent work by \citet{horvath2021clustering} proposed Wasserstein k-means (Wk-means) clustering as a powerful method to identify regimes in time series data, where they treated the case of one-dimensional asset returns in the financial context in particular.
Our paper builds upon that work and extends the method to multidimensional time series data.

Many traditional methods in the literature view regime shifts as abrupt structural changes in the coefficients of a model for the time series. Global break‑point search, as proposed in \cite{bai2003computation}, and the general‑to‑specific algorithm detailed in \cite{hendry2014empirical} attempt to locate these change points. Multivariate score‑driven extensions, such as in \cite{blazsek2024global}, follow the same idea of identifying structural change points, using Ward's hierarchical clustering technique in the space of the observed variables. The Wasserstein k‑means approach instead classifies regimes (and identifies changes between them) by analysing the behaviour of the \emph{distribution} of the time series itself throughout time, without relying on the idea of a model for the time series or structural changes in coefficients. This distributional approach marries the worlds of financial regime-switching and unsupervised clustering of distributions, as mentioned in \citet{horvath2021clustering}.

We begin by studying in detail the behaviour of the Wasserstein k-means clustering algorithm proposed in \citet{horvath2021clustering}, applied to one-dimensional (1d) time series data.
We extend the previous work by studying, in detail, the dynamics of the clustering algorithm and how varying the hyperparameters impacts the performance over different random initialisations.
We compute simple metrics that we find to be useful in identifying high-quality clusterings, which is especially important when ground-truth labels for regimes do not exist.

We then extend the technique of Wasserstein k-means clustering to multidimensional time series data by approximating the multidimensional Wasserstein distance via a sliced Wasserstein distance, as introduced by \cite{rabin2012wasserstein}. 
We call the resulting method `sliced Wasserstein k-means (sWk-means) clustering' and apply the method to the problem of automated regime classification in multidimensional time series.

We demonstrate the validity and effectiveness of the sWk-means method by applying it to synthetic multidimensional time series data. Again, we study how the hyperparameters impact the performance of the clustering algorithm.
Finally, we show that our method is able to identify market regimes in real multidimensional financial time series effectively and efficiently, using publicly available FX spot rate data as a case study.
We conclude with remarks about some limitations of our approach and potential complementary or alternative approaches.

\section{Materials and methods}

In this section, we begin by outlining the analytical framework employed in our study, including the Wasserstein metric.
We then introduce the concept of the \emph{sliced} Wasserstein distance as an approximation to the full Wasserstein distance in multiple dimensions, and explain how this concept allows us to formulate a sliced Wasserstein k-means (sWk-means) method that we use to cluster regimes in multidimensional time series. In formulating our sWk-means algorithm, we sidestep the need to find the full multidimensional Wasserstein barycentre by using fixed projection directions throughout the algorithm. This makes the method simple to implement and computationally efficient. In the remainder of this section, we follow closely the notation and definitions employed in \cite{horvath2021clustering} and \cite{kidger2019deep}.

\subsection{Data streams and empirical distributions}
\label{sec:data_streams}

We begin by defining a space $\mathcal{X}$ that our elementary data inhabits. In this paper, we take $\mathcal{X} = \mathbb{R}^d$; the case considered in practice in \citeauthor{horvath2021clustering} was $\mathcal{X} = \mathbb{R}$.

The fundamental object of interest in the analysis of time series is a stream of data $S \in \mathcal{S}(\mathcal{X})$, where the set $\mathcal{S}$ of streams of data over $\mathcal{X}$ is given by
\begin{equation}
\mathcal{S}(\mathcal{X}) = \{ \mathbf{x} = (x_1, \ldots, x_n) \, : \, x_i \in \mathcal{X}, n \in \mathbb{N} \},
\end{equation}
where $n$ is the length of a stream of data. 
In the setting of finance, a stream of length $N$,
\begin{equation}
S = (s_1, \ldots, s_N) \in \mathcal{S}(\mathbb{R}^d),
\end{equation}
might be a $d$-dimensional price path realised by a set of $d$ assets as a function of time~$t$, discretely observed. 

We can define transformations $r^S$ of the stream $S$, where $r^S \in \mathcal{S}(\mathcal{X}')$. For example, one such transformation corresponds to taking the log-returns of a price path $S$
\begin{equation}
r_i^S = \log(s_{i+1}) - \log(s_i),
\label{eq:log_returns} 
\end{equation}
in which case, $\mathcal{X}' = \mathcal{X} = \mathbb{R}^d$. We can standardise these coordinate-wise, without loss of generality (w.l.o.g.),~such that $\mathbb{E}(r^{S}) = 0$ and $\mathrm{Var}(r^{S}) = 1$.

We can further define a so-called lifting transformation $\ell$ that maps the set of streams $\mathcal{S}(\mathcal{X})$ into the set of streams of streams
\begin{equation}
\ell = (\ell^1, \ldots, \ell^M) \, : \, \mathcal{S}(\mathcal{X}) \rightarrow \mathcal{S}(\mathcal{S}(\mathcal{X})), 
\end{equation}
\textls[-25]{with $M > 1$.
One choice for the function $\ell$, proposed in the context of path signatures in \cite{kidger2019deep}}, is a sliding window transformation $\ell = \ell_{h_1 h_2}$ with window size $h_1$ and sliding window offset parameter (or `lifting size') $h_2$,
\begin{equation}
\ell^{m}(\mathbf{x}) = (x_{1+h_2(m-1)},x_{1+h_2(m-1)+1}, \ldots, x_{1+h_2(m-1)+h_1-1}) \quad \text{for} \; m = 1, \ldots, M,
\label{eq:lifting_sliding} 
\end{equation}
where $M \equiv \lfloor \frac{N - (h_1 - h_2)}{h_2} \rfloor$ is the maximum number of partitions that can be extracted from the stream of length $N$.
A stream of $M$ streams (or equivalently, a stream of sequences) can be obtained in this manner by applying $\ell$ to $r^S$. The lifting transformation of Equation~\eqref{eq:lifting_sliding} is illustrated pictorially in Figure~\ref{fig:lifting_transformation} for $h_1 = 6$ and $h_2 = 2$.

\begin{figure}[h]
	\begin{center}
\resizebox{\columnwidth}{!}
{
\begin{tikzpicture}[scale=2.5]

\normalsize
	\node at (0,0) {$\bullet$};
\node at (1,0) {$\bullet$};
\node at (2,0) {$\bullet$};
\node at (3,0) {$\bullet$};
\node at (4,0) {$\bullet$};
\node at (5,0) {$\bullet$};
\node at (6,0) {$\bullet$};
\node at (7,0) {$\bullet$};

	\node at (0,0.25) {{\large$x_1$}};
\node at (1,0.25) {{\large$x_2$}};
\node at (2,0.25) {{\large$x_3$}};
\node at (3,0.25) {{\large$x_4$}};
\node at (4,0.25) {{\large$x_5$}};
\node at (5,0.25) {{\large$x_6$}};
\node at (6,0.25) {{\large$x_7$}};
\node at (7,0.25) {{\large$x_8$}};

\draw [-, rounded corners] (0,-0.2) -- (0,-0.25) -- (5,-0.25) -- (5,-0.2);
	\node at (-1,-0.5) {$\bm{m=1}$};
	\node at (0,-0.5) {$x_{1+h_2(m-1) + i} = x_{1 + 0 + 0}$};
\node at (1,-0.5) {$x_{1 + 0 + 1}$};
\node at (2,-0.5) {$x_{1 + 0 + 2}$};
\node at (3,-0.5) {$x_{1 + 0 + 3}$};
\node at (4,-0.5) {$x_{1 + 0 + 4}$};
\node at (5,-0.5) {$x_{1 + 0 + 5}$};

\newcommand\Y3{-1}
\draw [-, rounded corners] (2,-0.95) -- (2,-1) -- (7,-1) -- (7,-0.95);
\draw [-, dashed] (2,-0.91) -- (2,-0.15);
\draw [-, dashed] (7,-0.91) -- (7,-0.15);
	\node at (-1,-1.25) {$\bm{m=2}$};
	\node at (2,-1.25) {$x_{1+h_2(m-1) + j} =x_{1 + 2 + 0}$};
\node at (3,-1.25) {$x_{1 + 2 + 1}$};
\node at (4,-1.25) {$x_{1 + 2 + 2}$};
\node at (5,-1.25) {$x_{1 + 2 + 3}$};
\node at (6,-1.25) {$x_{1 + 2 + 4}$};
\node at (7,-1.25) {$x_{1 + 2 + 5}$};

\draw [-, rounded corners] (4,-1.7) -- (4,-1.75) -- (7,-1.75);
\draw [-, dashed] (4,-1.66) -- (4,-0.15);
	\node at (-1,-2) {$\bm{m=3}$};
	\node at (4,-2) {$x_{1+h_2(m-1) + k} =x_{1 + 4 + 0}$};
\node at (5,-2) {$x_{1 + 4 + 1}$};
\node at (6,-2) {$x_{1 + 4 + 2}$};
\node at (7,-2) {$\ldots$};

	\node at (-1,-2.75) {$\bm{\ldots}$};


\end{tikzpicture}
}
		\caption[Pictorial illustration of the lifting transformation]{Pictorial description of the sliding window lifting transformation ${\ell^{m}(\mathbf{x}) = (x_{1+h_2(m-1)},x_{1+h_2(m-1)+1}, \ldots, x_{1+h_2(m-1)+h_1-1})} $ for $h_1 = 6$ and $h_2 = 2$; $m = 1,2,3$. Here, $h_1$ is the number of points in each sequence and $h_2$ is the separation between neighbouring sequences. Each stream (or sequence) generated by the lifting process defines an empirical measure  $\mu_m \equiv \frac{1}{h_1} \sum_i \delta_{\ell^m(\mathbf{x})_i}$.
}
	\label{fig:lifting_transformation}

	\end{center}
\end{figure}

To each stream (or sequence) generated by the lifting process of Equation~\eqref{eq:lifting_sliding}, \mbox{$\ell^m(\mathbf{x}) = \{\ell^m(\mathbf{x})_i : i = 1,\ldots,h_1\}$}, we can associate the discrete uniform distribution which defines an empirical measure 
\begin{equation}
\mu_m \equiv \frac{1}{h_1} \sum_i \delta_{\ell^m(\mathbf{x})_i},
\end{equation}
where $\delta_x$ is the Dirac delta at $x$, thus defining a family of such empirical measures
\begin{equation}
\mathcal{K} = \{ \mu_m\}_{1 \leq m \leq M},
\label{eq:empirical_measures}
\end{equation}
where $\mu_m \in \mathcal{P}_p(\mathbb{R}^d)$ for $m = 1,\ldots,M$, with $\mathcal{P}_p(\mathbb{R}^d)$ being the space of probability measures on $\mathbb{R}^d$ with finite $p^\text{th}$ moment.

It is the family of measures $\mathcal{K}$ defined by Equation~\eqref{eq:empirical_measures} that we wish to cluster, in the hope that the cluster to which each sequence is ascribed (via its empirical measure) will correspond to a certain regime characterised by some typical behaviour of the time series. If we take each measure $\mu_m$ in the family of measures $\mathcal{K}$ as corresponding to a point in some space, then we wish to achieve a clustering of the set of points in $\mathcal{K}$. A natural candidate for such a task is the k-means clustering algorithm, which is an unsupervised statistical learning algorithm. Note that, in order to define a k-means clustering algorithm over a set of points, we require notions of (i) the distance between pairs of points and (ii) a way of aggregating or averaging over a collection of points. For the case where the points to cluster are empirical distributions, these notions are naturally provided by the Wasserstein metric in the specific form of (i) the Wasserstein distance $\mathcal{W}_p$ and (ii) the Wasserstein barycentre $\bar{\mu}^{\mathcal{W}_p}$. The specific choice of the Wasserstein metric is motivated in more detail in Section~1.2 of \cite{horvath2021clustering}. This choice leads naturally to the Wasserstein k-means (Wk-means) algorithm as proposed by \citeauthor{horvath2021clustering} (2024) and employed in the restricted setting of $d=1$ in that paper.
Before summarising the Wasserstein k-means (Wk-means) algorithm, we introduce the Wasserstein metric, including the notions of Wasserstein distance and barycentre.

\subsection{The Wasserstein metric}
\label{sec:wkmeans_w-metric}

In this section, we introduce the Wasserstein metric, including the notions of Wasserstein distance and Wasserstein barycentre. We detail how the Wasserstein distance and barycentre can be computed efficiently when $d=1$ and introduce the notion of the sliced Wasserstein distance for $d>1$.

\subsubsection{The Wasserstein distance $\mathcal{W}_p$} 

Assume we have two probability measures $\mu$ and $\nu$ on $\mathcal{X} = \mathbb{R}^d$. Then, the $p$-Wasserstein distance between $\mu$ and $\nu$ is defined as the following infimum over the joint distributions $(X,Y)$ of $d$-dimensional random vectors $X$ and $Y$:
\begin{equation}
\mathcal{W}_p(\mu,\nu) = \inf_{\substack{(X,Y) \\ X \sim \mu  \\ Y \sim \nu}} (\mathbb{E} ||X - Y||^p)^{1/p},
\label{eqn:w-distance-defn}
\end{equation}
where $|| \cdot ||$ denotes some chosen norm on $\mathbb{R}^d$, $p \geq 1$, and $ X \sim \mu, Y \sim \nu$ indicates that $X$ and $Y$ are distributed according to $\mu$ and $\nu$, respectively (see \cite{panaretos2018statistical}, \cite{peyre2019computational} and \cite{stromme2020wasserstein}).
The joint distribution $(X,Y)$ satisfying Equation~\eqref{eqn:w-distance-defn} can be viewed as the optimal transport plan between $\mu$ and $\nu$ in the Kantorovich-type problem, i.e.,~the transport plan that minimises the effort required to reconfigure a mass distribution $\mu$ into the distribution $\nu$, where the effort required to move a unit of mass from position $x$ to position $y$ is given by $||x-y||^p$.
The $p$-Wasserstein distance exists for measures on $\mathcal{X}$ with finite $p^\mathrm{th}$ moment, a space that is denoted $\mathcal{P}
_p(\mathcal{X})$.
Note that an equivalent `analytic' definition of the Wasserstein distance is also commonly used.
The Wasserstein distance satisfies the axioms of a distance (see for example, Section~6 of \cite{villani2009optimal}, which includes an interesting discussion regarding the history of the Wasserstein distance and its name).

\subsubsection{The Wasserstein barycentre $\bar{\mu}^{\mathcal{W}_p}$} 

Suppose we have a family of probability measures $\mathcal{K}$ in a space $\mathcal{X}$ such as in Equation~\eqref{eq:empirical_measures}. Then, the Wasserstein barycentre  $\bar{\mu}^{\mathcal{W}_p}$ of $\mathcal{K}$  is that measure which minimises the total Wasserstein distance to the members of $\mathcal{K}$, that is,
\begin{equation}
\bar{\mu}^{\mathcal{W}_p} = \argmin_{\nu \in \mathcal{P}_p(\mathcal{X})} \sum_{\mu_m \in \mathcal{K}} \mathcal{W}_p(\nu, \mu_m)
\label{eq:w-barycentre}
\end{equation}
(see, for example, Definition~2.3 in \citeauthor{horvath2021clustering} (2024)). The existence of such a barycentre for the Wasserstein metric, which is easily computed for $d=1$, is one of the advantages of using the Wasserstein metric to formulate a k-means clustering algorithm to cluster the family of empirical measures $\mathcal{K}$. Presently, we introduce simple representations of the Wasserstein distance and Wasserstein barycentre in the case that $d=1$.

\subsubsection{$d=1$}

In the particular context of empirical distributions $\mu,\nu \in \mathcal{P}_p(\mathbb{R})$ with equal numbers of \textls[-20]{atoms $N$, $\{\mu_i\}_{1 \leq i \leq N}$, and $\{\nu_i\}_{1 \leq i \leq N}$, we have a particularly simple representation of the Wasserstein distance as follows:}
\begin{equation}
\mathcal{W}^p_p(\mu, \nu) = \frac{1}{N} \sum_{i=1}^N | \mu_i^* - \nu_i^* |^p,
\label{eq:w-distance-1d}
\end{equation}
where $\{\mu_i^*\}_{1 \leq i \leq N}$ and $\{\nu_i^*\}_{1 \leq i \leq N}$ are \emph{ordered} sequences corresponding to the atoms of $\mu$ and $\nu$, respectively  (see, for example, Proposition 2.6 in \cite{horvath2021clustering}, and Lemma~4.2 in \cite{bobkov2019one}). 
As a consequence of this representation, once the sequences  $\mu$ and $\nu$ are ordered (which need be done only once), the $p$-Wasserstein distance separating them can be computed efficiently, in an amount of time scaling linearly with the length of the sequences $N$.

In the same context of empirical distributions $\mu,\nu \in \mathcal{P}_p(\mathbb{R})$ with equal numbers of atoms $N$, the Wasserstein barycentre also has a simple representation that can be calculated efficiently. Concretely, given a family of $M$ empirical distributions \mbox{$\mathcal{K} = \{ \mu_m\}_{1 \leq m \leq M}$}, the Wasserstein barycentre $\bar{\mu}^{\mathcal{W}_p}$ of  $\mathcal{K}$ is given by
\begin{equation}
\bar{\mu}^{\mathcal{W}_p} = (\bar{\mu}_1, \ldots, \bar{\mu}_N),
\label{eq:w-barycentre-1d} 
\end{equation}
where, for $p=1$,
\begin{equation}
\bar{\mu}^{\mathcal{W}_p}_j = \text{Median}(\mu^*_{1,j}, \ldots, \mu^*_{M,j}) \quad \text{for}  \; j=1, \ldots, N \quad (p=1),
\label{eq:w-barycentre-1d-p1} 
\end{equation}
and, for $p=2$,
\begin{equation}
\bar{\mu}^{\mathcal{W}_p}_j = \text{Mean}(\mu^*_{1,j}, \ldots, \mu^*_{M,j}) \quad \text{for}  \; j=1, \ldots, N \quad (p=2)
\label{eq:w-barycentre-1d-p2} 
\end{equation}
(see, for example, \cite{peyre2019computational}, \cite{you2022wasserstein}, and the discussion of Fr\'echet means in \cite{bobkov2019one}).

\subsubsection{$d>1$}

When $d>1$, we do not have recourse to the simple representations to calculate the Wasserstein distance and Wasserstein barycentre that are given by Equations~\eqref{eq:w-distance-1d} and~\eqref{eq:w-barycentre-1d}, respectively. However, it is possible to approximate the full Wasserstein distance and barycentre via $d=1$  distances and barycentres corresponding to \emph{projections} of the full distributions. This formulation leads to the notions of the \emph{sliced} Wasserstein distance and barycentre, which we introduce below.

Note that it would also be possible to use optimisation methods to compute the Wasserstein barycentre without going via the sliced representation, especially when the number of dimensions is small. The usefulness of the sliced formalism, in our case, lies in being able to sidestep the task of finding the full \textls[25]{multidimensional barycentre, instead performing the clustering in the space of 1d projected measures, as we set out in more detail below. This also makes our implementation particularly computationally efficient.}

\paragraph{The sliced Wasserstein distance $\overline{\mathcal{W}}_p$}

Given two distributions in $d>1$, it is possible to approximate the full Wasserstein distance between the distributions as an integral of $d=1$ distances between projections of the full distributions. This is called the \emph{sliced} Wasserstein distance.

More explicitly, given an empirical measure $\mu \in \mathcal{P}_p(\mathbb{R}^d)$, we can define a \emph{projected} empirical measure $\mu'(\theta) \in \mathcal{P}_p(\mathbb{R})$, given by
\begin{equation}
\mu'(\theta) = \frac{1}{N} \sum_i \delta_{x'_i},
\label{eq:projected-distribution}
\end{equation}
where
\begin{equation}
x'_i = \langle x_i, \theta \rangle
\label{eq:projection}
\end{equation}
is the projection of $x_i$ along a vector $\theta \in \mathbb{S}^{d-1}$, with $\mathbb{S}^{d-1}$ the unit sphere in $d$ dimensions.
Then, the \emph{sliced} Wasserstein distance can be written as the integral
\begin{equation}
\overline{\mathcal{W}}_p(\mu, \nu) =\int_{\theta \in \mathbb{S}^{d-1}} \mathcal{W}_p(\mu'(\theta), \nu'(\theta)) d \theta.
\label{eq:sliced-w-distance}
\end{equation}
The terms $\mathcal{W}_p(\mu'(\theta), \nu'(\theta))$ in the sliced Wasserstein distance of Equation~\eqref{eq:sliced-w-distance} are 1d Wasserstein distances, which we know how to calculate efficiently from Equation~\eqref{eq:w-distance-1d}. 
In practice, the integral can be approximated as a sum over a finite number $L$ of projections,
\begin{equation}
\overline{\mathcal{W}}_p(\mu, \nu) \simeq \frac{1}{L} \sum_{l=1}^L \mathcal{W}_p(\mu'(\theta^l), \nu'(\theta^l)),
\label{eq:sliced-w-distance-approx}
\end{equation}
where the $\{ \theta^l \}$ are chosen from $\mathbb{S}^{d-1}$ -- in practice, a grid can be used, or the vectors can be randomly sampled via Monte Carlo, which results in advantageous scaling with dimensionality. As such, it is possible to approximate the full Wasserstein distance between the distributions $\mu$ and $\nu$ as a sum of $d=1$ distances between projections of the distributions $\mu'$ and $\nu'$ which we can calculate straightforwardly via Equation~\eqref{eq:w-distance-1d}.

\paragraph{The sliced Wasserstein barycentre $\bar{\mu}^{\overline{\mathcal{W}}_p}$}

As set out in Section~\ref{sec:data_streams}, in order to formulate a k-means clustering algorithm over a set of points we require notions of (i) the distance between points, and (ii) a way of averaging over sets of points, i.e.,~a barycentre.
For multidimensional time series, the sliced Wasserstein distance of Equation~\eqref{eq:sliced-w-distance} provides us with an efficient way of obtaining (i). 
For the barycentre, we can appeal to the notion of the sliced Wasserstein distance introduced above to define a sliced version of the Wasserstein barycentre, defined following Equation~\eqref{eq:w-barycentre} as the minimiser
\begin{equation}
\bar{\mu}^{\overline{\mathcal{W}}_p} = \argmin_{\nu \in \mathcal{P}_p(\mathcal{X})} \sum_{\mu_m \in \mathcal{K}} \overline{\mathcal{W}}_p^p(\nu, \mu_m),
\label{eq:sliced-w-barycentre}
\end{equation}
where $\overline{\mathcal{W}}_p(\mu, \nu)$ is the sliced Wasserstein distance given by Equation~\eqref{eq:sliced-w-distance}. The (multidimensional) sliced Wasserstein barycentre can then be found, in principle, from Equation~\eqref{eq:sliced-w-barycentre}, for example, via numerical optimisation such as in \cite{rabin2012wasserstein} or by backprojecting 1d barycentres of the projected distributions found using Equation~\eqref{eq:w-barycentre-1d} via the inverse Radon transform, such as in \cite{bonneel2013sliced} and \cite{bonneel2015sliced}. 
In practice, calculating the sliced Wasserstein barycentre via one of these methods is relatively computationally expensive, at least compared with the calculation of 1d barycentres using Equation~\eqref{eq:w-barycentre-1d}. For this reason, in our sWk-means algorithm, we will opt to use a set of fixed projection directions $\{\theta^l\}$ which define a grid on $\mathbb{S}^{d-1}$. This choice allows us to avoid computing the multidimensional (sliced) Wasserstein barycentre altogether and instead cluster the multidimensional distributions in the space of projected distributions. As an additional benefit, with this approach, the projected distributions need only be calculated and ordered once, which makes the method computationally efficient.

With these concepts in hand, we turn to detailing the sWk-means method that we use to cluster regimes in multidimensional time series.

\subsection{The sWk-means method}
\label{sec:swkmeans-method}

In this section, we detail the algorithm we use to cluster the family $\mathcal{K} \subset \mathcal{P}_p(\mathbb{R}^d)$ of empirical measures, obtained from the data stream $S \in \mathcal{S}(\mathbb{R}^d)$, using the k-means method along with the sliced Wasserstein distance and barycentre introduced in the previous section. An exposition of the k-means clustering algorithm in a classical setting can be found in~\cite{hartigan1975clustering}.

Given a dataset $S \in \mathcal{S}(\mathbb{R}^d)$,  we begin by applying a transformation $r^S$, which, in our case, consists of computing the (log) returns given by Equation~\eqref{eq:log_returns}, which we can standardise coordinate-wise w.l.o.g.,~such that $\mathbb{E}(r^{S_j}) = 0$ and $\mathrm{Var}(r^{S_j}) = 1$ for $j = 1, \ldots, d$. We then apply the lifting transformation $\ell(r^S)$ where $\ell$ is given by Equation~\eqref{eq:lifting_sliding}, which produces a family $\mathcal{K}$ of $M$ empirical distributions,  $\mathcal{K} = \{ \mu_j\}_{1 \leq j \leq M}$. At this stage, each empirical distribution $\mu_j$ is $d$-dimensional. We then choose a set of $L$ fixed vectors \mbox{$\{ \theta^l : l = 1, \ldots, L \}$} which define a grid on $\mathbb{S}^{d-1}$. For each of these vectors  $\theta^l$, we compute the projected distributions $ \{ \mu'_j (\theta^l) \}_{1 \leq j \leq M}$ via Equations~\eqref{eq:projected-distribution} and \eqref{eq:projection} for \mbox{$l=1,\ldots,L$}. 
The k-means algorithm begins by choosing the initial clusters by randomly picking $K$ distributions from $\mathcal{K}$ to use as the initial cluster centroids \mbox{$\{ \bar{\mu}_k : k =1,\ldots,K \}$}. The centroids $ \bar{\mu}_k \in \mathcal{P}_p(\mathbb{R}^d) $ are defined by their projections along $\{ \theta^l \}$, $\{ \bar{\mu}'_k(\theta^l) : l=1,\ldots, L \}$, where $ \bar{\mu}'_k(\theta^l) \in \mathcal{P}_p(\mathbb{R}) $. The projected distributions need only be computed and ordered once. 
We then perform the clustering. To do so, we iterate over all points (distributions) $\{ \mu_j : j = 1,\ldots,M \}$ and assign each point to a cluster~$\mathcal{C}_k$ based on the nearest centroid $ \bar{\mu}_k$ with respect to the sliced Wasserstein distance $\overline{\mathcal{W}}_p$ computed from Equation~\eqref{eq:sliced-w-distance-approx}.
We then update the centroid $\bar{\mu}_k$ as the sliced Wasserstein barycentre relative to $\mathcal{C}_k$ by updating the centroid projection $\bar{\mu}'_k(\theta^l)$ as the barycentre of the projected distributions belonging to cluster $\mathcal{C}_k$, \mbox{$\{ \mu'_j(\theta^l) : \mu_j \in \mathcal{C}_k\}$}, for each $\theta^l$, using Equation~\eqref{eq:w-barycentre-1d}. We iterate this procedure until convergence, which is defined in terms of the following criterion that determines when the cluster centroids have stopped moving within some tolerance $\epsilon$:
\begin{equation}
\sum_k \overline{\mathcal{W}}_p(\bar{\mu}^{n}_k, \bar{\mu}^{n-1}_k) < \epsilon,
\label{eq:stopping-criterion}
\end{equation}
where $\bar{\mu}_k^{n}$ denotes a centroid obtained after some iteration step $n$. We use a tolerance $\epsilon = 10^{-6}$ but the precise value has little effect on the convergence of the algorithm.

The sWk-means algorithm is summarised in Algorithm~\ref{alg:swkmeans}. The sWk-means algorithm can be \textls[20]{compared with the Wk-means algorithm, which is summarised in Algorithm~\ref{alg:wkmeans}, following \citeauthor{horvath2021clustering} (2024).}

\begin{algorithm}[t!]
 \KwResult{$K$ clusters}
\textbf{calculate} $\ell(r^S)$ given $S \in \mathcal{S}(\mathbb{R}^d)$\;
\textbf{define} family of empirical distributions $\mathcal{K} = \{ \mu_j\}_{1 \leq j \leq M}$ where $\mu_j \in \mathcal{P}_p(\mathbb{R}^d)$\;
\textbf{define} projection directions $\theta^l, l=1, \ldots, L$\;
\textbf{calculate} projected distributions $ \{ \mu'_j (\theta^l) \}_{1 \leq j \leq M}$, where $\mu'_j (\theta^l) \in \mathcal{P}_p(\mathbb{R}) $ is obtained from $\mu_j$ via projection on $\theta^l$\;
\textbf{initialise} centroids $ \bar{\mu}_k, k=1,\ldots, K $ by sampling $K$ times from $\mathcal{K}$,  where the centroids are defined by their projections $ \bar{\mu}'_k(\theta^l) \in \mathcal{P}_p(\mathbb{R}) $ for $l=1,\ldots,L$\;
 \While{convergence\_criterion $>$ tolerance}{
\ForEach{$\mu_j$}{
  \textbf{assign} to cluster $\mathcal{C}_k$ according to the closest centroid $\bar{\mu}_k$ with respect to (wrt) $\overline{\mathcal{W}}_p$ for $k = 1, \ldots, K$\;
   }
 \ForEach{$\theta^l$}{
   \textbf{update} centroid projection $\bar{\mu}'_k(\theta^l) $ as the Wasserstein barycentre of projected distributions $\mu'_j(\theta^l)$ in cluster $\mathcal{C}_k$;
}
   \textbf{calculate} convergence\_criterion\;
  }
 \caption{sWk-means algorithm}
\label{alg:swkmeans}
\end{algorithm}

\begin{algorithm}[h!]
 \KwResult{$K$ clusters}
\textbf{calculate} $\ell(r^S)$ given $S \in \mathcal{S}(\mathbb{R})$\;
\textbf{define} family of empirical distributions $\mathcal{K} = \{ \mu_j\}_{1 \leq j \leq M}$\;
\textbf{initialise} centroids $\bar{\mu}_k, k = 1,\ldots,K$ by sampling $K$ times from $\mathcal{K}$\;
 \While{convergence\_criterion $>$ tolerance}{
 \ForEach{$\mu_j$}{
  \textbf{assign} to cluster $\mathcal{C}_k$ according to the closest centroid $\bar{\mu}_k$ wrt $\mathcal{W}_p$ for $k = 1, \ldots, K$\;
   }
   \textbf{update} centroid $\bar{\mu}_k$ as the Wasserstein barycentre of distributions $\mu_j$ in cluster $\mathcal{C}_k$\;
   \textbf{calculate} convergence\_criterion\;
  }
 \caption{Wk-means algorithm
\citep{horvath2021clustering}
}
\label{alg:wkmeans}
\end{algorithm}

\newpage
\section{Results}


In Section~\ref{sec:1d_ts}, we study the behaviour of the Wk-means clustering algorithm in detail, using synthetic 1d time series data. We investigate the dynamics of the algorithm and the effect of varying the hyperparameters, showing how these impact the performance of the clustering algorithm for different random initialisations. Compared with the more limited investigation in Section~3.4 of \citeauthor{horvath2021clustering}, our systematic analysis considers the interplay between $h_1$ and $h_2$ in detail, as well as the effects of varying $h_2$ and how this depends on the dataset size. We also identify and compute two metrics that are useful in identifying high-quality clusterings.

In Sections~\ref{sec:2d_ts} and \ref{sec:3d_ts-synthetic}, we study the behaviour of the sWk-means clustering algorithm applied to synthetic multidimensional time series data. We demonstrate that the sWk-means method is able to identify the regimes in the data before again investigating the effect of varying the hyperparameters.

Finally in Section~\ref{sec:2d_ts-real-data} we apply the sWk-means algorithm to real-world multidimensional time series data, using publicly available FX spot rate data as a case study.

Throughout the remainder of the paper, we set \mbox{$p=1$}.

\subsection{1d time series data: Dynamics and performance of the Wk-means algorithm}
\label{sec:1d_ts}

In this section, we  study the behaviour of the Wk-means clustering algorithm for 1d time series data. In Section~\ref{sec:1d_ts-synthetic_data_generation}, we detail how we construct synthetic 1d data; we then give an example of clustering results on the synthetic 1d data in Section~\ref{sec:1d_ts-example}. In Section \ref{sec:1d_ts-dynamics} we use the synthetic data to study the dynamics of the algorithm. We study several metrics throughout the iterations of the algorithm and suggest how these metrics can be used to identify good clusterings. Next, in Section~\ref{sec:1d_ts-accuracy}, we quantify the accuracy of the clustering algorithm for different combinations of the window size parameter~$h_1$ and the lifting size parameter~$h_2$. We demonstrate how the quality of the clustering results can depend on the amount of data available, and this leads us to suggest a way of optimising the clustering results in low-data environments.

\subsubsection{The 1d synthetic data generation method}
\label{sec:1d_ts-synthetic_data_generation}

In this section, we detail how we construct synthetic 1d time series data, allowing us to study the accuracy of the algorithm for different combinations of the window size parameter~$h_1$ and the lifting size parameter~$h_2$.

We use a synthetic data generation method that is analogous to one of those employed in \cite{horvath2021clustering}, namely geometric Brownian motion with regimes corresponding to `bullish' and `bearish' parameters. We begin by summarising this synthetic data generation method.

Geometric Brownian motion paths $S_t / S_{t-1} =\exp(r_t^S)$ with parameters $\Theta$ are constructed from log returns $r_t$ distributed according to
\begin{equation} 
r_t^S \sim N \left( (\mu - \sigma^2/2)dt, \sigma^2 dt \right),
\label{eq:gbm-eqn}
\end{equation}
where $N(\cdot)$ is the normal distribution and $\Theta$ represents the parameters of the geometric Brownian motion, namely the annualised mean (log) return $\mu$ and the annualised standard deviation of (log) returns $\sigma$:
\begin{equation}
\Theta \equiv (\mu, \sigma).
\end{equation}

We consider a period of 20 years, with 252 days in a year and 7 (hourly) observations per day. As such, the time increment $dt$ in Equation~\eqref{eq:gbm-eqn} is given by $dt = 1/(252 \times 7)$ and there are $35,280$ data points in total. We generate paths corresponding to `bullish' parameters $\Theta_\text{bull}$ everywhere apart from 10~half-year periods with `bearish' parameters $\Theta_\text{bear}$, where the starting points of the `bearish' periods are randomly chosen subject to the periods being non-overlapping. We use the following parameter values for the `bullish' and `bearish' regimes:
\begin{align}
\Theta_\text{bull} & = (0.02, 0.2), \\
\Theta_\text{bear} & = (-0.02, 0.3),
\label{eq:1d_bull_bear_params}
\end{align}
which are the same as those used by \citeauthor{horvath2021clustering}.

\begin{figure}[h!]
\centering
\includegraphics[width=\linewidth]{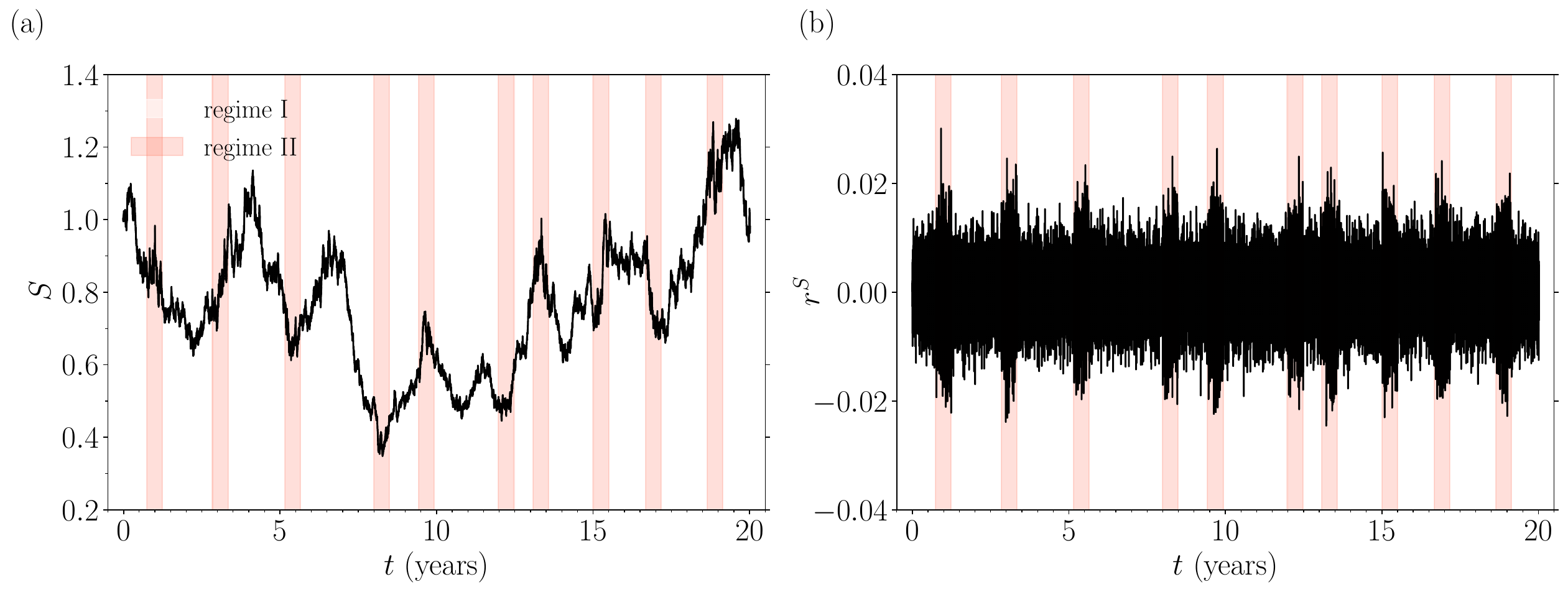}
\caption{Synthetic 1d data containing two regimes. (a) The time series $S(t)$,  with the majority regimes (I) corresponding to `bullish' parameters $\Theta_\mathrm{bull}$ and minority regimes (II) corresponding to `bearish' parameters $\Theta_\mathrm{bear}$ indicated. (b) The corresponding log returns $r^S$. There are ${20 \times 252 \times 7 = 35,280}$ data points.} 
\label{fig:ts1-synthetic-figure}
\end{figure}

An example of a path $S(t)$ constructed in this manner, with the majority `bullish' regimes (I) and minority `bearish' regimes (II) indicated, is illustrated in Figure~\ref{fig:ts1-synthetic-figure}(a), along with the corresponding log returns $r^S$ in Figure~\ref{fig:ts1-synthetic-figure}(b).

The fact that we use synthetic data with explicitly constructed regimes allows us to compute the accuracy of the clustering results. We define the accuracy of the clustering in a slightly simpler way than \citeauthor{horvath2021clustering}, which we now outline.

Recall that the Wk-means algorithm clusters sequences (via their empirical distributions). Each point in the time series can belong to more than one sequence, depending on the parameters of the lifting transformation $h_1$ and $h_2$, and thus each time point can be assigned more than one label. We determine the overall label (regime) of a point in the time series via a simple majority voting mechanism, with cases where there is an equal split resolved in favour of the prevailing label (regime). This is a conservative choice, since it enforces the principle that regimes switch only when a majority of the labels switch in favour of a new regime. 
Once having benefitted from the data augmentation and probabilistic classification that the lifting transformation provides, the majority voting mechanism returns all the points in the time series to an equal footing.

Then, mathematically, with each data point $r^S_{t_i}$ assigned a single label $\hat{y}_{t_i}$, the total accuracy (TA) achieved by a clustering $\mathcal{C} = \{ \hat{y}_{t_i}\}_{1\leq i \leq N}$ within a given partition $\tilde{t}$ of the time series is given by
\begin{equation}
\text{TA}(\mathcal{C},\tilde{t}) =  \frac{\sum_{t_i \in \tilde{t}} \mathbb{I}_{\hat{y}_{t_i} = y_{t_i}}}{\sum_{t_i \in \tilde{t}} (\mathbb{I}_{\hat{y}_{t_i} = y_{t_i}} + \mathbb{I}_{\hat{y}_{t_i} \neq y_{t_i}} )},
\label{eq:accuracy}
\end{equation}
where $ y_{t_i}$ is the true label of data point $r^S_{t_i}$. We can also define the accuracy within a given regime $k$ by taking $\tilde{t} = \{ t_i : y_{t_i} = k \}$.

\subsubsection{Clustering example}
\label{sec:1d_ts-example}

\begin{figure}[h!]
\centering
\includegraphics[width=\linewidth]{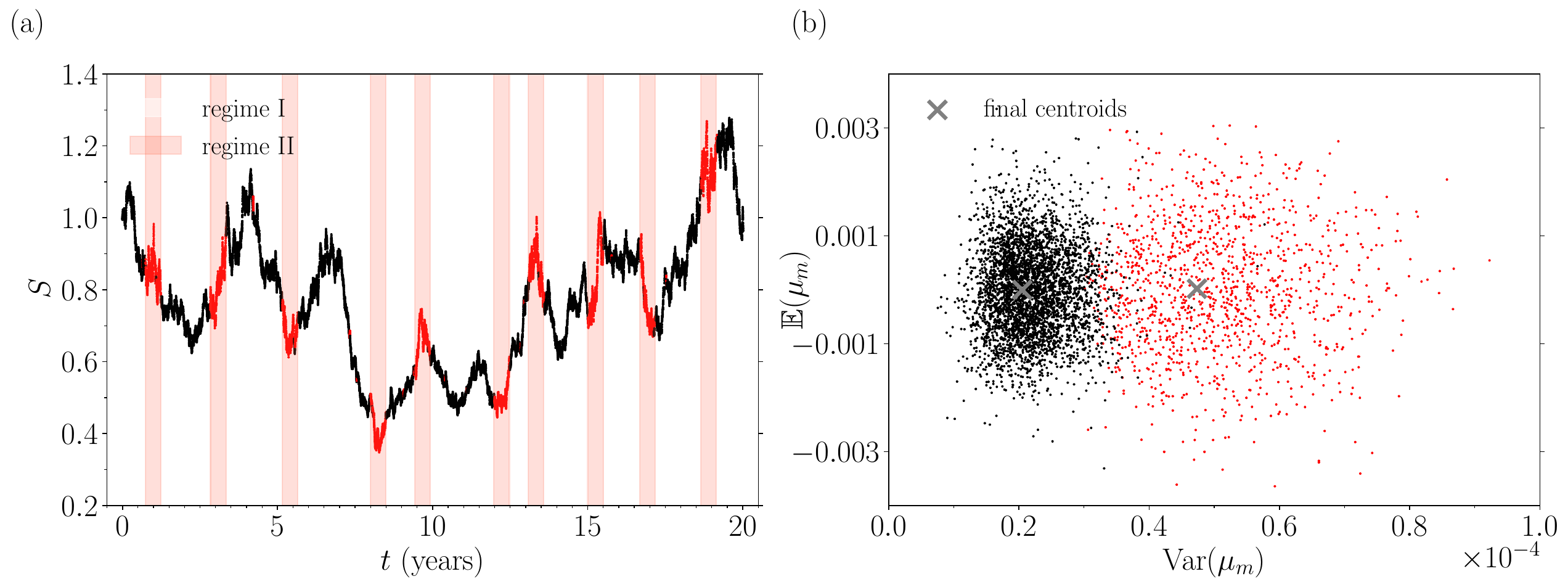}
\caption{Results of the Wk-means clustering algorithm applied to the synthetic 1d data shown in Figure~\ref{fig:ts1-synthetic-figure}. (a) Clustering results for the time series $S(t)$. Each point in the time series is coloured according to its assigned cluster. (b) Clustering results for the distributions $\mu_m \in \mathcal{K}$ in mean-variance ($\text{Var}(\mu_m)$-$\mathbb{E}(\mu_m)$) space. Each point is coloured according to its assigned cluster. The window size is  $h_1 = 35$ and the lifting size is $h_2 = 7 \, (20\%)$.} 
\label{fig:ts1-synthetic-1-run}
\end{figure}

The results of the Wk-means clustering algorithm applied to the synthetic 1d data plotted in Figure~\ref{fig:ts1-synthetic-figure} are shown in Figure~\ref{fig:ts1-synthetic-1-run}, using a window size $h_1 = 35$, a lifting size $h_2 = 7 \, (20\%)$\footnote{Note that here and throughout the remainder of the paper, in addition to its numerical value, we will also specify the value of $h_2$ in terms of a percentage of $h_1$.}, and $K=2$ clusters. Each point in the time series is coloured according to its assigned cluster, using majority voting for points that belong to more than one sequence, as discussed in Section~\ref{sec:1d_ts-synthetic_data_generation}. In Figure~\ref{fig:ts1-synthetic-1-run}(b), each empirical distribution $\mu_m \in \mathcal{K}$ is plotted in mean-variance ($\text{Var}(\mu_m)$-$\mathbb{E}(\mu_m)$)  space, again with each point coloured according to its assigned cluster, alongside the locations of the final cluster centroids.

Having illustrated a clustering example, in the next section, we explore the dynamics of the algorithm applied to the 1d synthetic data for different random initialisations.

\subsubsection{Dynamics}
\label{sec:1d_ts-dynamics}

At the start of the Wk-means algorithm detailed in Algorithm~\ref{alg:wkmeans} with $K$ clusters, centroids are initialised by sampling randomly $K$ times from $\mathcal{K}$. Just as with traditional applications of the k-means clustering algorithm, the algorithm may (and likely will) converge to different final states, depending on the initialisation of the centroids' locations.
\textls[-25]{In this section, we study some aspects of the dynamics of the algorithm for different random initialisations, which gives us an insight into the performance of the algorithm.}

We compute and plot the following quantities during the evolution of the clustering algorithm for different random initialisations:
\begin{itemize}
\item The mean squared point--centroid distance, given by
\begin{equation}
 \langle \mathcal{W}_p(\mu_i, \bar{\mu}_k)^2 \rangle_{k,i\in\mathcal{C}_k} := \frac{1}{K} \sum_{k} ||\mathcal{C}_k||^{-1} \sum_{i \in \mathcal{C}_k}  \mathcal{W}_p(\mu_i, \bar{\mu}_k)^2,
\label{eq:mean-sq-point-centroid-d}
\end{equation}
\noindent where $\bar{\mu}_k$ is the centroid of the cluster $\mathcal{C}_k$. This is essentially the same as the within-cluster variation discussed in \cite{horvath2021clustering}.

\begin{figure}[H]
\centering
\includegraphics[width=\linewidth]{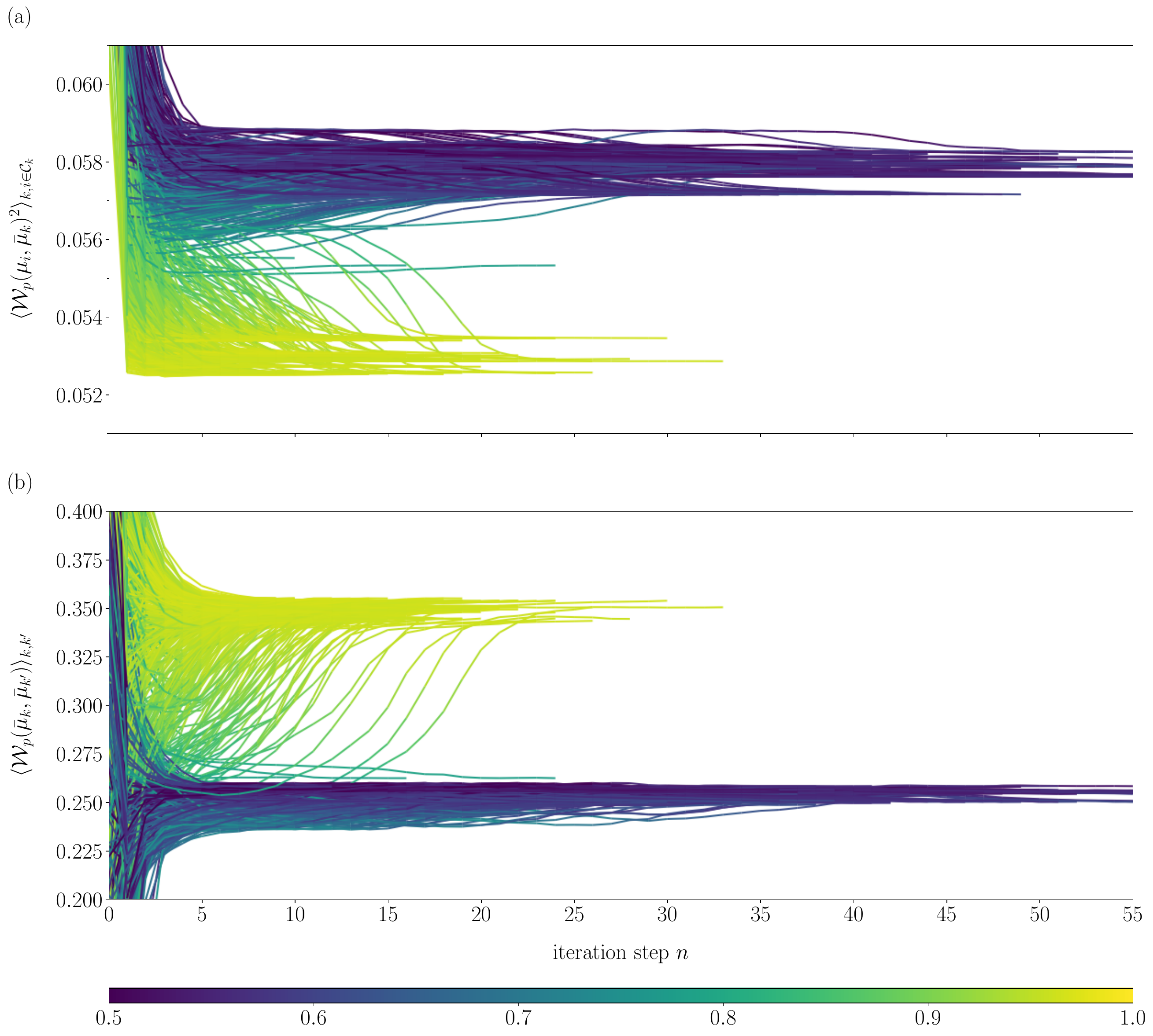}
\caption{Dynamics of the Wk-means clustering algorithm applied to the synthetic 1d data shown in Figure~\ref{fig:ts1-synthetic-figure}. (a) Mean squared point--centroid distance $ \langle \mathcal{W}_p(\mu_i, \bar{\mu}_k)^2 \rangle_{k,i\in\mathcal{C}_k} $ and (b)~mean centroid--centroid distance $  \langle \mathcal{W}_p(\bar{\mu}_k, \bar{\mu}_{k'}) \rangle_{k,k'} $ as a function of the algorithm iteration for different random initialisations. The paths are coloured according to the instantaneous total accuracy $\mathrm{TA}(\mathcal{C})$ computed during the evolution of the algorithm (see the colourbar). The two metrics are effective in differentiating between high- and low-accuracy clusterings. The window size is  $h_1 = 30$ and the lifting size is $h_2 = 9 \, (30\%)$.} 
\label{fig:1d_ts_dynamics}
\end{figure}

\item The mean centroid--centroid distance, given by
\begin{equation}
 \langle \mathcal{W}_p(\bar{\mu}_k, \bar{\mu}_{k'}) \rangle_{k,k'} := (C^K_2)^{-1} \sum_{k' \neq k} \mathcal{W}_p(\bar{\mu}_k, \bar{\mu}_{k'}),
\label{eq:mean-centroid-centroid-d}
\end{equation}
where $C^K_2 \equiv \binom{K}{2}$ is a combinatorial coefficient. This is similar to metrics such as the cluster separation or $c$-separation discussed in \cite{kanungo2002efficient} and \cite{dasgupta1999learning}, considering the pairwise separations of all the centroids.
\end{itemize}

In addition to the random initialisation of the centroid locations, for each clustering run, we introduce a random offset $0 \leq \delta \leq h_2-1$ to the lifting transformation in order to alleviate edge effects associated with the regime locations being fixed in the data (as they are in reality).

The results for $\langle \mathcal{W}_p(\mu_i, \bar{\mu}_k)^2 \rangle_{k,i\in\mathcal{C}_k}$ and $ \langle \mathcal{W}_p(\bar{\mu}_k, \bar{\mu}_{k'}) \rangle_{k,k'}$ as a function of the algorithm iteration when applied to the synthetic 1d data shown in Figure~\ref{fig:ts1-synthetic-figure}, for different random initialisations, can be seen in Figure~\ref{fig:1d_ts_dynamics}(a) and~(b), respectively, using a window size  $h_1 = 30$ and a lifting size $h_2 = 9 \, (30\%)$. The paths are coloured according to the instantaneous total accuracy $\mathrm{TA}(\mathcal{C})$ computed during the evolution of the \textls[20]{algorithm, with yellow corresponding to high accuracy, and blue corresponding to low accuracy (see the colourbar).} 

Two bands can be seen in the metrics, corresponding to high and low final values of $\langle \mathcal{W}_p(\mu_i, \bar{\mu}_k)^2 \rangle_{k,i\in\mathcal{C}_k}$ and $ \langle \mathcal{W}_p(\bar{\mu}_k, \bar{\mu}_{k'}) \rangle_{k,k'}$. Interestingly, the bands represent predominantly a single colour, meaning that these metrics can be used to track or determine the accuracy of the clusterings: Specifically, the paths with high final values of  $ \langle \mathcal{W}_p(\bar{\mu}_k, \bar{\mu}_{k'}) \rangle_{k,k'}$ and low final values of $\langle \mathcal{W}_p(\mu_i, \bar{\mu}_k)^2 \rangle_{k,i\in\mathcal{C}_k}$ tend to have high final accuracies (in yellow); conversely, the paths with low final values of  $ \langle \mathcal{W}_p(\bar{\mu}_k, \bar{\mu}_{k'}) \rangle_{k,k'}$ and high final values of $\langle \mathcal{W}_p(\mu_i, \bar{\mu}_k)^2 \rangle_{k,i\in\mathcal{C}_k}$ tend to have low final accuracies (in blue).
As such, the mean squared point--centroid distance $\langle \mathcal{W}_p(\mu_i, \bar{\mu}_k)^2 \rangle_{k,i\in\mathcal{C}_k}$ and  mean centroid--centroid distance $ \langle \mathcal{W}_p(\bar{\mu}_k, \bar{\mu}_{k'}) \rangle_{k,k'}$ can be used to differentiate between high- and low-accuracy clusterings, and we will use the latter later on in this paper as an easily computed, objective numerical metric to determine high-quality clusterings to retain and plot.

Having illustrated the dynamics of the algorithm through the prism of the metrics defined in Equations~\eqref{eq:mean-sq-point-centroid-d} and \eqref{eq:mean-centroid-centroid-d}, we proceed by outlining how the performance of the algorithm depends on the choice of hyperparameters.
\newpage

\subsubsection{Effect of hyperparameters on accuracy}
\label{sec:1d_ts-accuracy}

In this section, we study the effect of varying the different hyperparameters, particularly, the window size $h_1$ and window offset parameter $h_2$, on the accuracy of the clustering results for the synthetic 1d data shown in Figure~\ref{fig:ts1-synthetic-figure}.

We run $N_c = 1,000$ clusterings with different random initialisations; for each clustering $\mathcal{C}$, we compute the total accuracy $\mathrm{TA}(\mathcal{C})$. We can then compute the statistics of  $\mathrm{TA}(\mathcal{C})$ over the different clusterings including, for example, the average total accuracy $\overline{\mathrm{TA}} = \overline{\mathrm{TA}(\{ \mathcal{C} \})}$, in the form of the mean or median value.  Note that we use the same data (with fixed regime locations) for all $N_c$ clusterings in order to best reflect the fixed (historical) locations of regimes experienced in reality; as before, we use a random offset $0 \leq \delta \leq h_2-1$ to the lifting transformation for each clustering run.

In addition to the full 20-year synthetic dataset presented in Section~\ref{sec:1d_ts-synthetic_data_generation}~and~\ref{sec:1d_ts-example}, we also analyse reduced 2-year and 1-year datasets, containing $3,530$ and $1,765$ data points respectively, obtained simply by taking the first two (respectively, one) year(s) from the full 20-year dataset. These reduced datasets allow us to investigate the effect that `low-data' environments have on the accuracy of the clustering algorithm.

The results for the median and maximum values of $\mathrm{TA}$ over the $N_c = 1,000$ runs can be seen in Table~\ref{table:ts1}. We also show the accuracies of the clusterings corresponding to the maximum value of the mean centroid--centroid distance  $\langle \mathcal{W}_p(\bar{\mu}_k, \bar{\mu}_{k'}) \rangle_{k,k'} := (C^K_2)^{-1} \sum_{k' \neq k} \mathcal{W}_p(\bar{\mu}_k, \bar{\mu}_{k'})$ introduced and discussed in the preceding section.

\begin{table}[h!]
\centering
\caption{Effect of $h_1$ and $h_2$ parameters on the accuracy of the Wk-means clustering algorithm. Statistics for total accuracy $\mathrm{TA}(\mathcal{C})$ for $N_c = 1,000$ clustering runs using 1-year, 2-year, and 20-year subsets of the synthetic 1d data shown in Figure~\ref{fig:ts1-synthetic-figure}.}
\label{table:ts1}
\input{./tables/ts1_accuracy_matrix_k_2.tex}
\end{table}

As a broad trend, we observe that the median and maximum accuracies increase with increasing $h_1$: Larger window sizes correspond to a greater number of data points in the sequences, which allows the algorithm to better capture the differences between them (via their underlying distributions), being less susceptible to sampling bias and small-scale noise. Empirically, there appears to be a `critical' value of $h_1$ below which there is insufficient statistical information in the sequences for the algorithm to capture the salient distributional information, leading to few, if any, clusterings with acceptable accuracies ($h_1 < 20$ in this example). It is reasonable to assume that the numerical value of this `critical' value of $h_1$ is likely to depend on the particular dataset (and the underlying distributions) to which the method is applied, rather than being universal across different datasets. Another tradeoff to note is that increasing the value of $h_1$ decreases the sensitivity of the algorithm in detecting changes between regimes, which can be an important consideration, especially when using the algorithm in an online manner.

\begin{figure}[t!]
\centering
\includegraphics[width=0.7\linewidth]{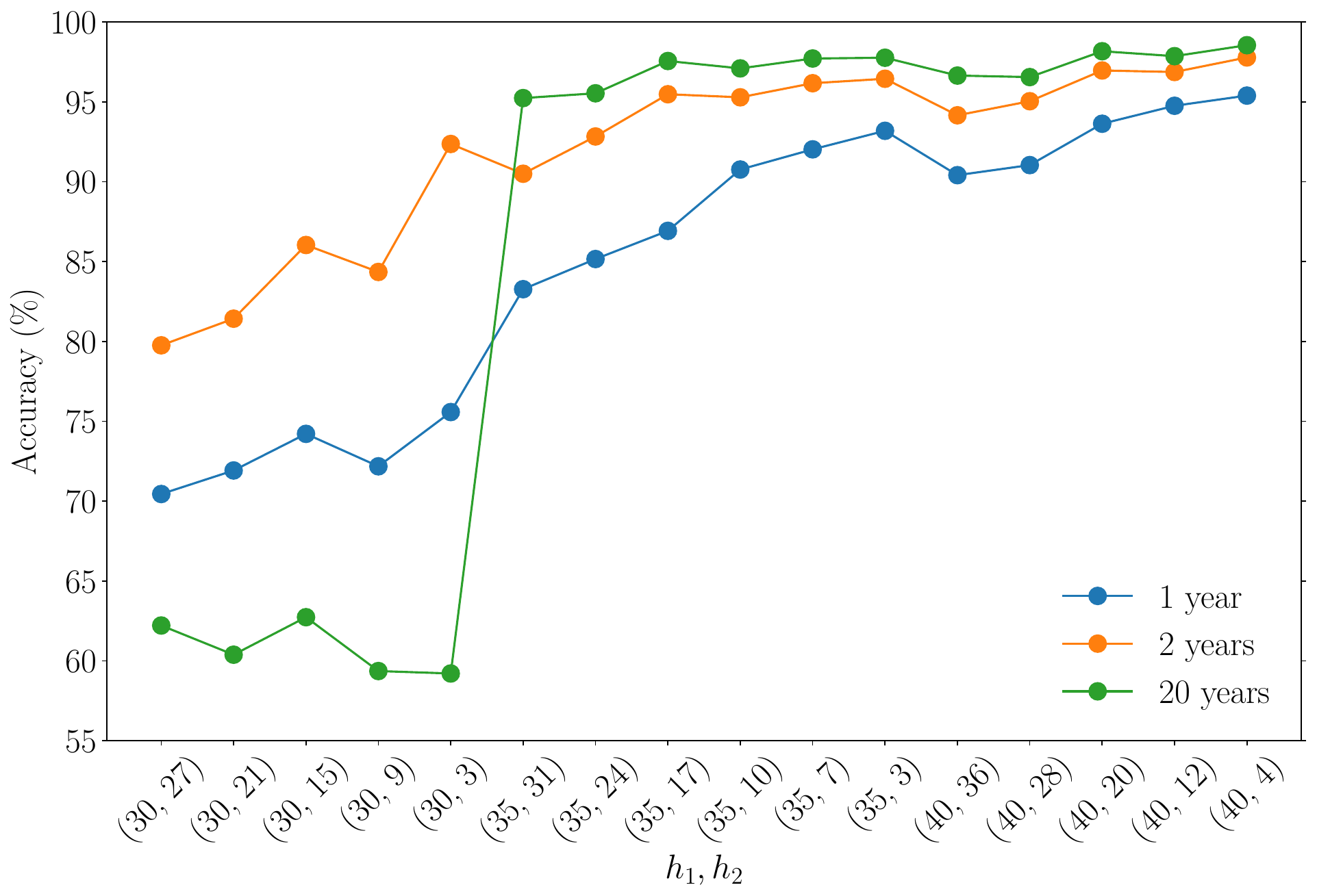}
\caption{Dependence of the average accuracy score $\overline{\mathrm{TA}}$ (median) computed from $N_c = 1,000$ clustering runs on the window and lifting size ($h_1, h_2$), using different amounts of synthetic 1d data.
The average accuracy $\overline{\mathrm{TA}}$ generally increases with decreasing $h_2$ due to the data augmentation effect associated with decreasing $h_2$. This effect is particularly pronounced for smaller datasets (2 years, 1 year). The average accuracy also generally increases with $h_1$.
} \label{fig:ts1-synthetic-accurcy_plot}
\end{figure}

As a secondary broad trend, we observe that decreasing $h_2$ (which, for our definition of $h_2$, corresponds to increasing the overlap between successive sequences) can again increase the median accuracies of the clusterings. In order to better illustrate this behaviour, in Figure~\ref{fig:ts1-synthetic-accurcy_plot}, we have plotted the dependence of the average (median) accuracy $\overline{\mathrm{TA}}$ from Table~\ref{table:ts1} on the value of $h_2$ for different (increasing) values of $h_1 \geq 30$. The average accuracy $\overline{\mathrm{TA}}$  displays a generally increasing trend as a function of decreasing $h_2$, with increasing steps in the background accuracy level as $h_1$ is itself increased. 
We note that decreasing $h_2$ has a particularly pronounced effect on the average accuracy for the 1- and 2-year datasets. We attribute this behaviour to the data augmentation effects associated with decreasing $h_2$, i.e.,~increasing the overlap between successive sequences, which generates a greater number of sequences used as an input to the clustering algorithm for the same underlying data. All else being equal, the k-means clustering algorithm is a data-hungry method, benefitting in terms of performance when supplied with more data, and decreasing $h_2$ allows us to achieve this, which can be particularly beneficial in small data environments, such as the reduced 1- and 2-year reduced datasets.

Regarding the more limited investigation in \citeauthor{horvath2021clustering}, Section~3.4, we agree with the statements regarding $h_1$ and $h_2$ therein and have provided results to support this. Choosing values for $h_1$ and $h_2$ in real-world settings depends on the objective and is more of an art than a science. Although accuracy generally increases with $h_1$ due to the increased distributional information available, regimes cannot be distinguished on timescales much smaller than $h_1$. Therefore, $h_1$ should not be significantly larger than the timescale of interest in a real-world study. There are fewer tradeoffs involved with $h_2$.

Having studied how the accuracy of the clustering results depends on the hyperparameters $h_1$ and $h_2$, we now turn to investigating the performance of our proposed sWk-means method applied to multidimensional time series data.

\subsection{2d time series data: the sWk-means algorithm}
\label{sec:2d_ts}

In this section, we study the behaviour of the sWk-means algorithm proposed in Section~\ref{sec:swkmeans-method}, applied to synthetic two-dimensional (2d) time series data. 

We begin in Section~\ref{sec:2d_ts-synthetic_data_generation} by constructing sets of synthetic 2d data containing either two or three regimes. Then in Section~\ref{sec:2d_ts-results}, we show that the sWk-means algorithm performs well on the synthetic data. In Section~\ref{sec:2d_ts-n-projs}, we explore the accuracy metrics using different sets of hyperparameters.

\subsubsection{The 2d synthetic data generation method}
\label{sec:2d_ts-synthetic_data_generation}

We generate synthetic data in a manner similar to that described in Section \ref{sec:1d_ts-synthetic_data_generation}, with the method extended to two dimensions. The synthetic data we generate contain either two or three regimes, where two of the regimes are characterised by log returns having joint distributions that are Gaussian with a given correlation $\rho$, and the third by a more complex (highly non-Gaussian) structure.

More specifically, when the synthetic data contain two regimes, we use a correlated 2d geometric Brownian motion $S_t / S_{t-1}  =\exp(r_t^S)$, where \mbox{$S_t = (S_t^{(1)}, S_t^{(2)})$}, and each regime (denoted I and II) is characterised by log returns \mbox{$r_t = (r_t^{S^{(1)}}, r_t^{S^{(2)}})$} having a joint distribution that is a correlated Gaussian corresponding to a given set of parameters $\Theta = (\mu, \sigma)$ from Equation~\eqref{eq:1d_bull_bear_params} and a correlation $\rho$. We use the same regime locations and number of data points as previously. To obtain the geometric Brownian motions $S^{(1),(2)}_t$, we first generate two independent sets of log returns $r^{S^{(1)}}_t$ and $r^{S^{(')}}_t$ with parameters $\Theta$; the log returns ${r}^{S^{(2)}}_t$ having a correlation $\rho$ with $r^{S^{(1)}}_t$ are then generated as follows:
\begin{equation} 
r_t^{S^{(2)}} = \rho r^{S^{(1)}}_t + \sqrt{1 - \rho^2} r^{S^{(')}}_t.
\label{eq:2d_bull_bear_returns_dist}
\end{equation}

When the synthetic data contain three regimes, two of the regimes correspond to geometric Brownian motions generated as just described, and the additional third regime (denoted III) corresponds to either an additional geometric Brownian motion, or a geometric Brownian motion-like process characterised by log returns having a joint distribution exhibiting a more complex, highly non-Gaussian structure that we describe as `moon-shaped', generated with the \verb+datasets.make_moons()+ function in the \texttt{sklearn} Python package \citep{scikit-learn}. We rotate and scale the moon-shaped distribution produced by this function to endow it with a given correlation $\rho$ and a mean and variance corresponding to a given set of parameters $\Theta$. We set the noise value to 5\% arbitrarily.

A summary of the parameters of our 2d synthetic datasets can be found in Table~\ref{tab:2d-data-summary}.

\begin{table}[th!]
\begin{center}

\caption{Summary of synthetic 2d dataset parameters.}
\setlength{\tabcolsep}{10mm}
\begin{tabular}{ c  c  c  c }
\hline 
 Type & Regime I & Regime II & Regime III \\
\hline 
 Type A & $\Theta_\text{bull}; \rho = + 1/2$ & $\Theta_\text{bear}; \rho = + 1/2$ & N/A \\  
 Type B & $\Theta_\text{bull}; \rho = + 1/2$ & $\Theta_\text{bull}; \rho = - 1/2$ & N/A \\
 Type C &  $\Theta_\text{bull}; \rho = + 1/2$ & $\Theta_\text{bear}; \rho = + 1/2$ & $\Theta_\text{bear}; \rho = - 1/2$ \\
 Type D &  $\Theta_\text{bull}; \rho = + 1/2$ & $\Theta_\text{bear}; \rho = + 1/2$ & $\Theta_\text{moon}; \rho = + 1/2$\\
\hline
\end{tabular}
\label{tab:2d-data-summary}
\end{center}
\end{table}

The first and second sets of synthetic 2d data (which we denote Types~A and B) contain two regimes.

\begin{figure}[th!]
\centering
\includegraphics[width=1.0\linewidth]{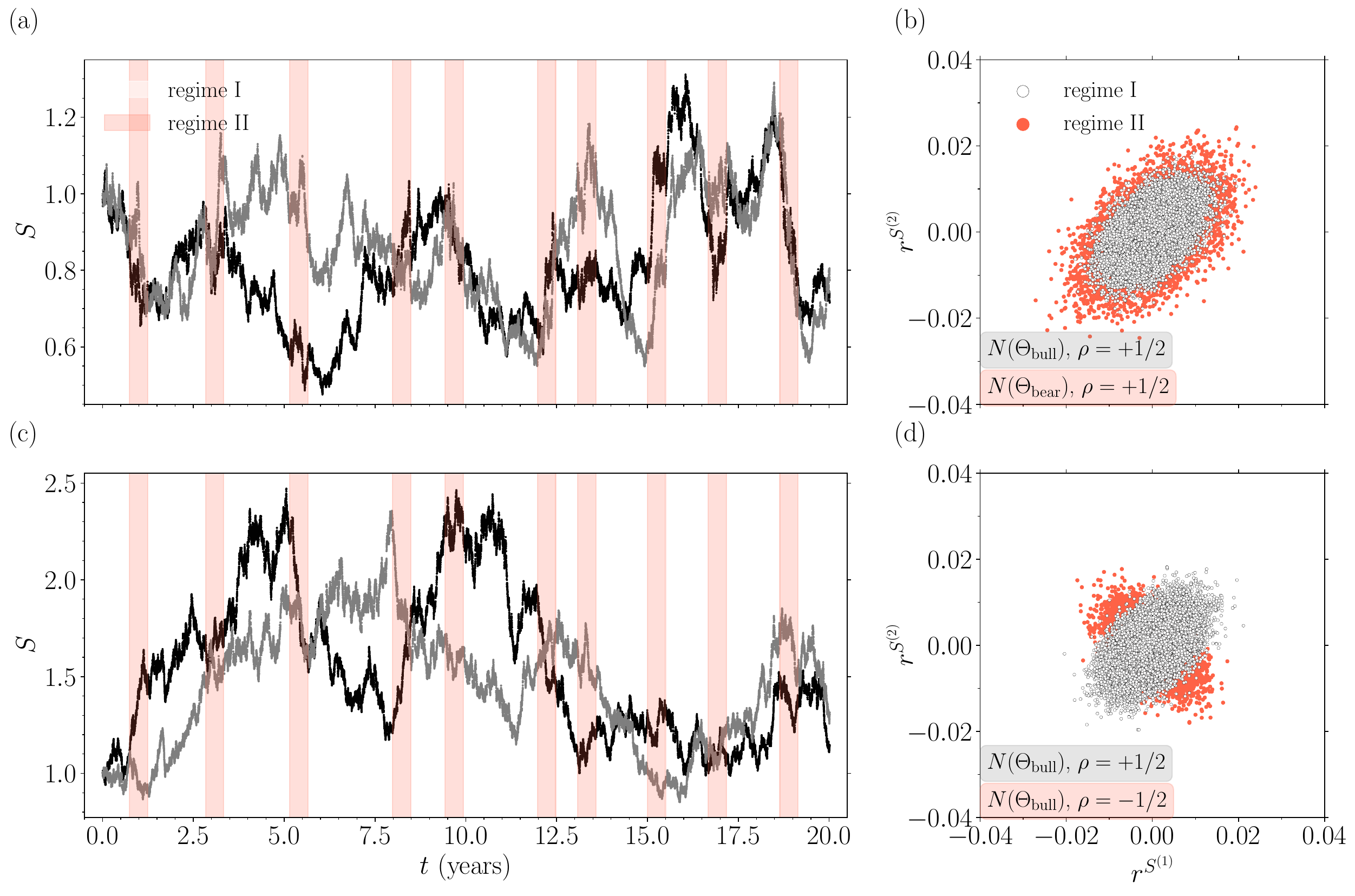}
\caption{Synthetic 2d time series data with two regimes. (a), (c)  The time series $S(t)$, with the majority (I) and minority (II) regimes indicated. (b), (d) The empirical distributions of log returns $r^S$ corresponding to (a) and~(c), respectively. There are \mbox{$20 \times 252 \times 7 = 35,280$} data points. 
The data in (a) and~(b) have regime~I corresponding to `bullish' parameters $\Theta_\mathrm{bull}$, regime~II corresponding to `bearish' parameters $\Theta_\mathrm{bear}$, and $\rho = +1/2$ for both regimes (Type~A).
The data in (c) and~(d) have regime~I and II both corresponding to `bullish' parameters $\Theta_\mathrm{bull}$, but regime~I having $\rho = +1/2$ and regime~II having $\rho = -1/2$ (Type~B).
The light-coloured points in the distributions in (b) and~(d) correspond to the majority regime (I) periods with no highlighting in (a) and~(c); the orange points correspond to the minority regime (II) periods highlighted in orange.
}
\label{fig:ts2-synthetic-figure-1}.
\end{figure}

In the first set of synthetic 2d data, the majority regime (I) has `bullish' parameters $\Theta_\textrm{bull}$ and the minority regime (II) has `bearish' parameters $\Theta_\textrm{bear}$, where the parameters $\Theta_\mathrm{bull}$ and $\Theta_\mathrm{bear}$ are given by Equation~\eqref{eq:1d_bull_bear_params},  and we set $\rho = +1/2$ for both regimes. We denote this type of synthetic data as Type~A. An example of such a 2d path $S(t) = (S_t^{(1)}, S_t^{(2)})$ can be seen in Figure~\ref{fig:ts2-synthetic-figure-1}(a), with the empirical distribution of returns  $r_t = (r_t^{S^{(1)}}, r_t^{S^{(2)}}) $ shown in Figure~\ref{fig:ts2-synthetic-figure-1}(b). Note that the light-coloured points in the distributions correspond to the majority regime~I periods with no highlighting in Figure~\ref{fig:ts2-synthetic-figure-1}(a); the orange points correspond to the minority regime~II (bearish) highlighted in orange in Figure~\ref{fig:ts2-synthetic-figure-1}(a). 

\begin{figure}[th!]
\centering
\includegraphics[width=\linewidth]{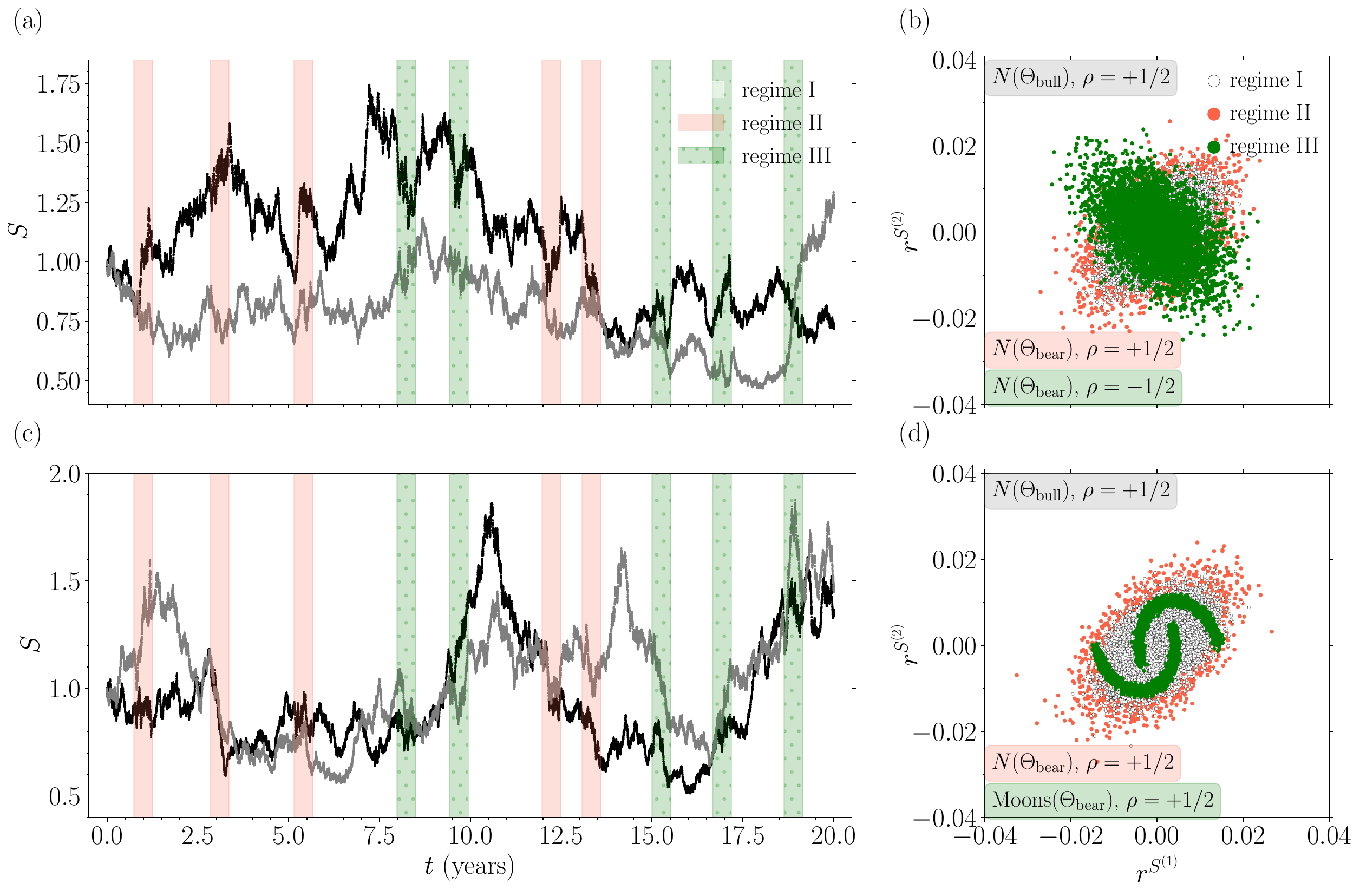}
\caption{Synthetic 2d time series data with three regimes. (a), (c)  The time series $S(t)$, with the majority (I) and minority (II and III) regimes indicated. (b), (d) The empirical distributions of log returns $r^S$ corresponding to (a) and~(c), respectively. There are \mbox{$20 \times 252 \times 7 = 35,280$} data points. 
The data in (a) and~(c) have regime~I corresponding to `bullish' parameters $\Theta_\mathrm{bull}$ with $\rho = +1/2$, regime~II corresponding to `bearish' parameters $\Theta_\mathrm{bear}$ with $\rho = +1/2$, and regime~III corresponding to `bearish' parameters $\Theta_\mathrm{bear}$ with $\rho = -1/2$ (Type~C).
The data in (b) and~(d) have regime~I corresponding to `bullish' parameters $\Theta_\mathrm{bull}$ with $\rho = +1/2$, regime~II corresponding to `bearish' parameters $\Theta_\mathrm{bear}$ with $\rho = +1/2$, and regime~III having a joint distribution characterised by `bearish' parameters $\Theta_\mathrm{bear}$ and $\rho = +1/2$, like regime II, but with a more complex, highly non-Gaussian `moon-shaped' structure (Type~D).
The light-coloured points in the distributions in (b) and~(d) correspond to the majority regime (I) periods with no highlighting in (a) and~(c); the orange and green points correspond to the minority regime (II and III) periods highlighted in orange and green respectively.
}
\label{fig:ts2-synthetic-figure-2}.
\end{figure}

In the second set of synthetic 2d data, both the majority and minority regimes (I and II) have `bullish' parameters  $\Theta_\textrm{bull}$, but we set $\rho = +1/2$ for regime~I and $\rho = -1/2$ for \mbox{regime II}. We denote this type of synthetic data as Type~B. An example of such a 2d path $S(t) = (S_t^{(1)}, S_t^{(2)})$,  along with the empirical distribution of returns, can be seen in Figure~\ref{fig:ts2-synthetic-figure-1}(c) and~(d).

\textls[-25]{The third and fourth sets of synthetic 2d data (which we denote as Types~C and D) contain three regimes.}

In our third set of synthetic 2d data, the majority regime (I) has `bullish' parameters $\Theta_\textrm{bull}$ and the second (minority) regime (II) has `bearish' parameters $\Theta_\textrm{bear}$, both with $\rho = +1/2$. Then, the third (minority) regime~(III) has `bearish' parameters $\Theta_\textrm{bear}$, and $\rho = -1/2$.  We denote this type of synthetic data as Type~C. An example of such a 2d path $S(t) = (S_t^{(1)}, S_t^{(2)})$, along with the empirical distribution of returns can be seen in Figure~\ref{fig:ts2-synthetic-figure-2}(a) and~(b). Again, the points in the empirical distributions in Panel~(b) are coloured according to the corresponding regimes in Panel~(a).

In the fourth set of synthetic 2d data, the majority regime (I) has `bullish' parameters $\Theta_\textrm{bull}$, and the second (minority) regime (II) has `bearish' parameters $\Theta_\textrm{bear}$, both with $\rho = +1/2$. Then, the third (minority) regime~(III) corresponds to a moon-shaped distribution with `bearish' parameters $\Theta_\textrm{bear}$, and correlation $\rho = +1/2$. As such, the mean, variance, and correlation of regime~III exactly match those of regime II; their joint distributions differ only in the more complex details of their structure.  We denote this type of synthetic data as Type~D. An example of such a 2d path $S(t) = (S_t^{(1)}, S_t^{(2)})$, along with the empirical distribution of returns, can be seen in Figure~\ref{fig:ts2-synthetic-figure-2}(c) and~(d).

Having described the 2d synthetic data, we now turn to discussing the results of the sWk-means clustering algorithm on these data.
\newpage

\begin{figure}[th!]
\centering
\includegraphics[width=1.0\linewidth]{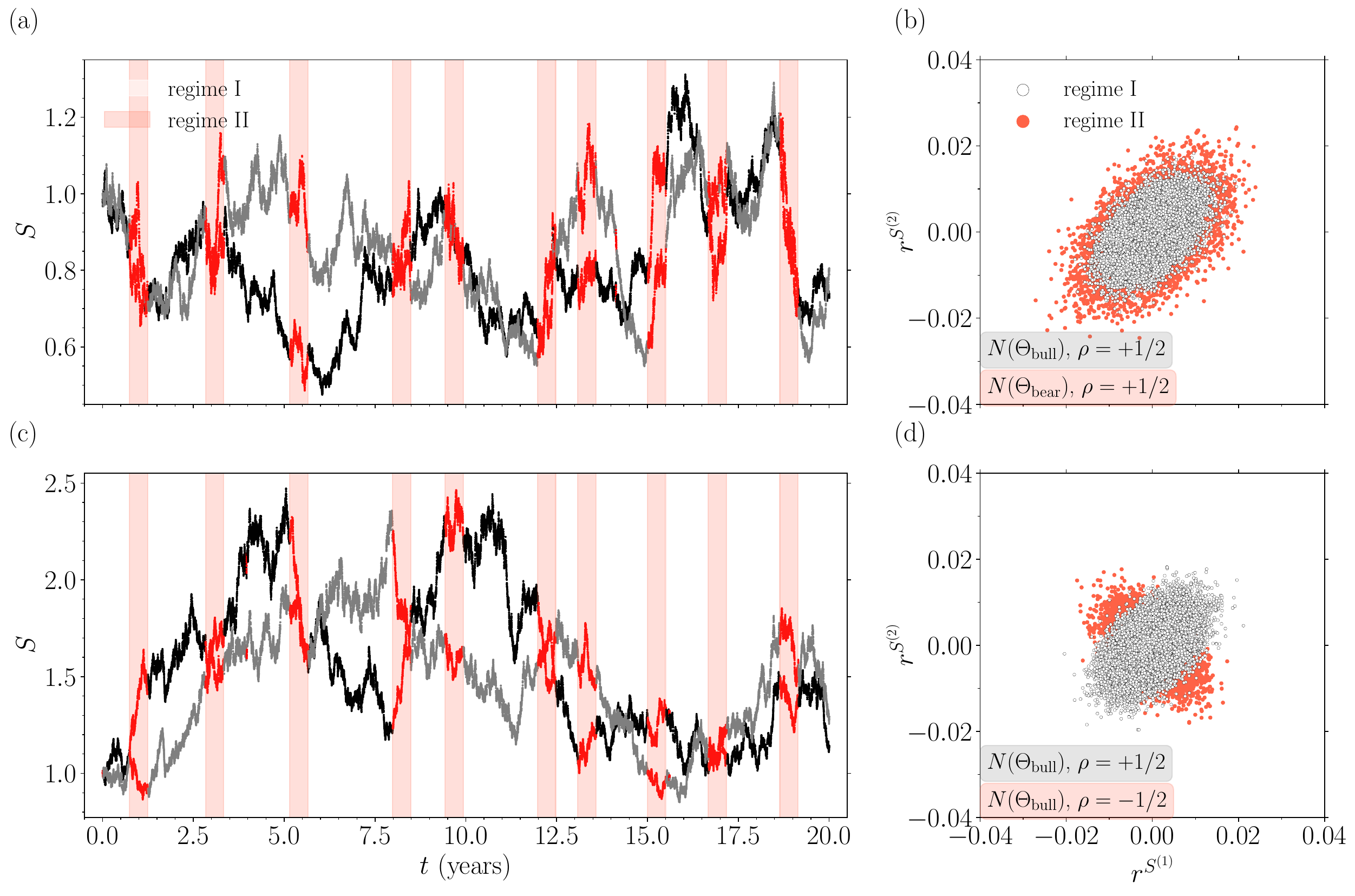}
\caption{Results of the sWk-means algorithm applied to the synthetic 2d time series data with two regimes shown in Figure~\ref{fig:ts2-synthetic-figure-1}. The colouration of the points in the time series reflects the cluster assigned by the algorithm.
The window size is $h_1=35$; the lifting size is \mbox{$h_2 = 7 \, (20\%)$}, the number of projections is  $L=9$, and the number of clusters is $K=2$.
The run with the largest final mean centroid--centroid distance $  \langle \mathcal{W}_p(\bar{\mu}_k, \bar{\mu}_{k'}) \rangle_{k,k'} $ from 100 independent runs with different random initialisations was chosen.
}
\label{fig:ts2-synthetic-figure-1-results}
\end{figure}

\subsubsection{Results}
\label{sec:2d_ts-results}

In this section we describe the results of the sWk-means clustering algorithm introduced in Section~\ref{sec:swkmeans-method} on the synthetic 2d time series data generated as described in the preceding section and illustrated in Figures~\ref{fig:ts2-synthetic-figure-1} and \ref{fig:ts2-synthetic-figure-2}.
For each set of data, we chose the run with the largest final mean centroid--centroid distance $ \langle \mathcal{W}_p(\bar{\mu}_k, \bar{\mu}_{k'}) \rangle_{k,k'} $ from 100 independent runs with different random initialisations. 
\newpage

Figure~\ref{fig:ts2-synthetic-figure-1-results} shows the results of the sWk-means clustering algorithm applied to the synthetic 2d data with two regimes, illustrated in Figure~\ref{fig:ts2-synthetic-figure-1}. The coloration of the points in the time series reflects the cluster assigned by the algorithm. Like the 1d case in Section~\ref{sec:1d_ts}, we use a window size of $h_1=35$ and a lifting size of $h_2 = 7 \, (20\%)$. The number of projections used is $L=9$, and the number of clusters is $K=2$.
As is clear from the plot, the sWk-means algorithm is very effective at clustering the two regimes in the data. 
We set out the numerical accuracy metrics in more detail in the next section.

\begin{figure}[th!]
\centering
\includegraphics[width=\linewidth]{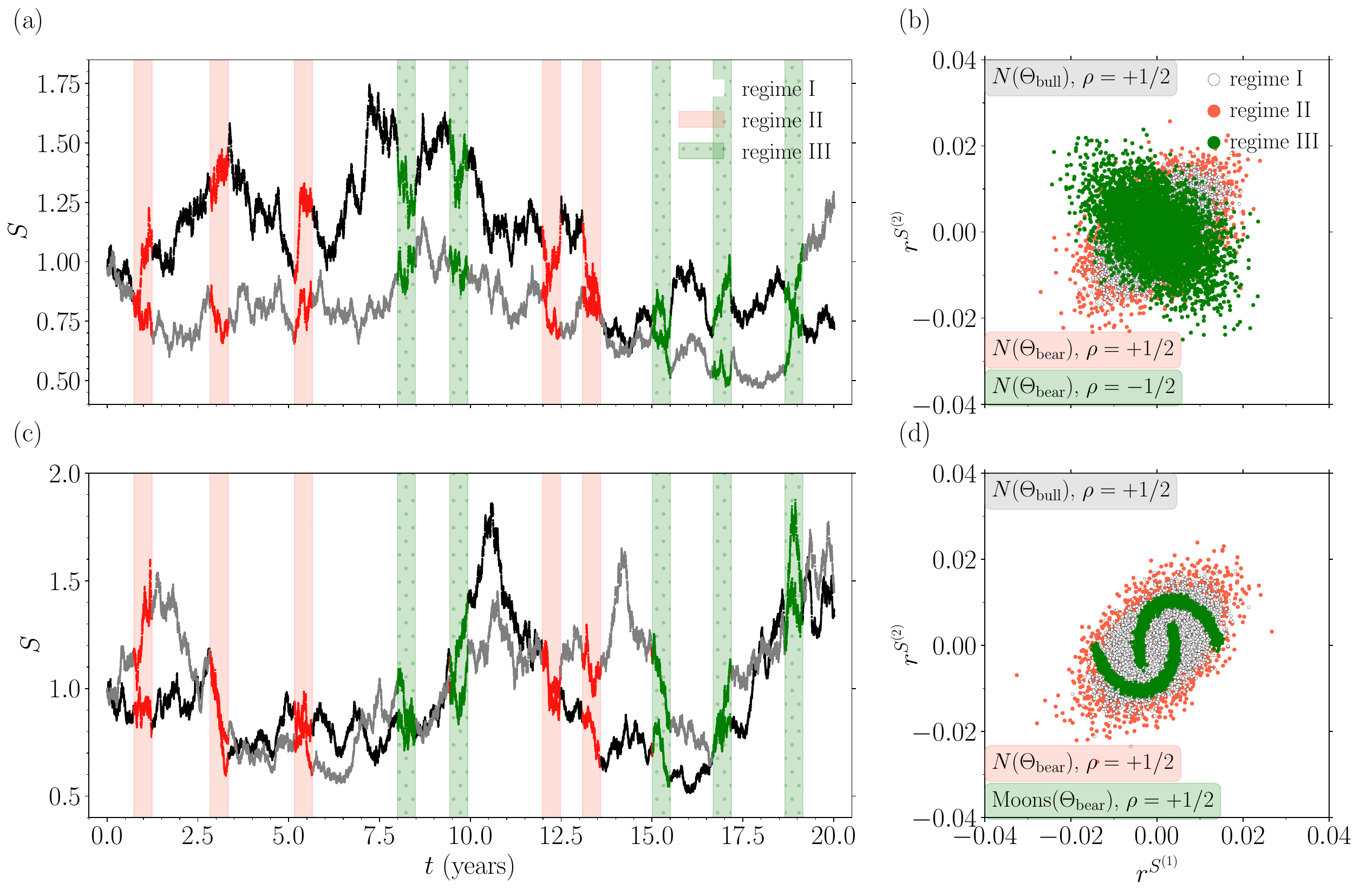}
\caption{Results of the sWk-means algorithm applied to the synthetic 2d time series data with three regimes shown in Figure~\ref{fig:ts2-synthetic-figure-2}. The colouration of the points in the time series reflects the cluster assigned by the algorithm.
The window size is $h_1=60$; the lifting size is $h_2 = 12 \, (20\%)$, the number of projections is  $L=9$, and the number of clusters is $K=3$.
The run with the largest final mean centroid--centroid distance $  \langle \mathcal{W}_p(\bar{\mu}_k, \bar{\mu}_{k'}) \rangle_{k,k'} $ from 100 independent runs with different random initialisations was chosen.
}
\label{fig:ts2-synthetic-figure-2-results}
\end{figure}

Figure~\ref{fig:ts2-synthetic-figure-2-results} shows the results of the sWk-means clustering algorithm applied to the synthetic 2d data with three regimes illustrated in Figure~\ref{fig:ts2-synthetic-figure-2}. Here, we have increased the window size to $h_1=60$ and the lifting size in proportion to $h_2=12 \, (20\%)$. Again, the number of projections used is $L=9$, and the number of clusters is $K=2$.
We can see that the sWk-means algorithm is very effective at clustering the three regimes in the data. Note that in the last set of 2d synthetic data, regime~II and regime~III have exactly the same means and variances ($\Theta$); they also have identical correlations $\rho$. Therefore, \emph{a priori}, it is not trivial for the algorithm to differentiate these two regimes; to do so, it must rely on finer details of the distributions and is successful nonetheless. Note that, if $K=2$ clusters are used for this dataset instead of $K=3$, the algorithm groups regimes II and III into the same cluster. Since these regimes can reasonably be considered the most similar, this is reassuring.

We now move on to considering the accuracy metrics computed over a set of independent runs with different random initialisations. 

\subsubsection{Accuracy metrics}
\label{sec:2d_ts-n-projs}

In this section, we study the effect of varying the different hyperparameters on the accuracy of the sWk-means algorithm. In addition to varying the window size $h_1$ (and the window offset parameter $h_2$ as a fixed 20\% fraction of $h_1$), we also vary the number of projections $L$ that are used. We run $N_c = 100$ clusterings with different random initialisations; for each clustering $\mathcal{C}$, we compute the total accuracy $\mathrm{TA}(\mathcal{C})$. We can then compute the statistics of  $\mathrm{TA}(\mathcal{C})$ over the different clusterings including the average (median) total accuracy $\overline{\mathrm{TA}} = \overline{\mathrm{TA}(\{ \mathcal{C} \})}$.

The results for the median and maximum values of $\mathrm{TA}$ over the $N_c = 100$ runs for the Type~A and Type~B data containing two regimes (illustrated in Figure~\ref{fig:ts2-synthetic-figure-1}) can be seen in Table~\ref{table:ts2_k2}.

\begin{table}[t!]
\centering
\caption{Effect of window size $h_1$ and number of projections $L$ on the accuracy of the sWk-means clustering algorithm for synthetic 2d data containing two regimes (Types A and B, shown in Figure~\ref{fig:ts2-synthetic-figure-1}). The median and maximum values of the accuracy metric $\mathrm{TA}(\mathcal{C})$ over $N_c = 100$ clustering runs are shown, along with the accuracy of the clustering identified via the maximum of the mean centroid--centroid distance $ \langle \mathcal{W}_p(\bar{\mu}_k, \bar{\mu}_{k'}) \rangle_{k,k'} $.}
\label{table:ts2_k2}
{\tiny
\input{./tables/ts2_k2_new_acc.tex}}
\end{table}

\begin{table}[h!]
\centering
\caption{Effect of window size $h_1$ and number of projections $L$ on the accuracy of the sWk-means clustering algorithm for synthetic 2d data containing three
 regimes (Types C and D, shown in Figure~\ref{fig:ts2-synthetic-figure-2}). The median and maximum values of the accuracy metric $\mathrm{TA}(\mathcal{C})$ over $N_c = 100$ clustering runs are shown, along with the accuracy of the clustering identified via the maximum of the mean centroid--centroid distance  $ \langle \mathcal{W}_p(\bar{\mu}_k, \bar{\mu}_{k'}) \rangle_{k,k'} $.}
\label{table:ts2_k3}
{\tiny
\input{./tables/ts2_k3_new_acc.tex}}
\end{table}

The regimes in the synthetic 2d data of Type~A have the same correlations, but different marginal distributions. The regimes in the Type~B synthetic data have the same marginal distributions, but different correlations. Accordingly, as can be seen in Table~\ref{table:ts2_k2}, a minimum number of four projections is required to cluster the Type~B synthetic data since using only two projections captures only the marginal distributions, which are the same in both regimes. However, for the Type~A synthetic data, two projections are sufficient to cluster the regimes, since the regimes differ in their marginal distributions. Increasing the value of $L$ increases the average accuracy. This is particularly visible for the Type~B data, which have more subtle differences between the regimes (i.e., identical marginals but different correlations). Analogous to the 1d case, increasing the value of $h_1$ increases the accuracy of the clusterings, and there is some `critical' value of $h_1$, below which few, if any, clusterings have acceptable accuracies, due to each sequence containing insufficient information to capture the details of the different distributions. For the Type~B data, there are some intermediate values of $h_1$, where the maximum of the metric $ \langle \mathcal{W}_p(\bar{\mu}_k, \bar{\mu}_{k'}) \rangle_{k,k'} $ is unable to identify the most accurate clusterings; however, this appears to be a transient effect that disappears when $h_1$ is increased further.

The results for the median and maximum values of $\mathrm{TA}$ over the $N_c = 100$ runs for the Type~C and Type~D data containing three regimes (illustrated in Figure~\ref{fig:ts2-synthetic-figure-2}) can be seen in Table~\ref{table:ts2_k3}.
\newpage
For this synthetic data, the results for the average (median) accuracy $\overline{\mathrm{TA}}$ are poor for all the hyperparameter combinations, though slightly better for the Type~D data. That being said, the clusterings identified via the maximum of the mean centroid--centroid distance metric  $ \langle \mathcal{W}_p(\bar{\mu}_k, \bar{\mu}_{k'}) \rangle_{k,k'} $ tend to have accuracies very close to the maximum accuracy, again demonstrating the utility of this metric. For the Type~D data, which contains two regimes that have the same means, variances, and correlations (regimes II and III) that are thus hard to differentiate, there are some discrepancies between the maximum accuracies and those identified via the maximum of  $ \langle \mathcal{W}_p(\bar{\mu}_k, \bar{\mu}_{k'}) \rangle_{k,k'} $; however, these discrepancies disappear with a sufficiently large value of $h_1$. Again, we attribute this behaviour to the requirement that the sequences contain enough information for the algorithm to be effective in differentiating regimes~II and III.

\subsection{3d time series data}
\label{sec:3d_ts-synthetic}

Having shown in the preceding section that our algorithm performs well for synthetic 2d time series data, in this section, we illustrate the application to three-dimensional (3d) data generated in a similar manner. We restrict our study to datasets containing only two regimes; the algorithm can deal with more regimes straightforwardly. We begin in Section~\ref{sec:3d_ts-synthetic_data_generation} by outlining the synthetic datasets that we construct before detailing the results of the algorithm in Section~\ref{sec:3d_ts-synthetic_results}.

\subsubsection{3d synthetic data generation method}
\label{sec:3d_ts-synthetic_data_generation}

To generate the 3d synthetic time series data, we sample log returns from a 3d multivariate normal distribution as follows:
\begin{equation}
r_t^S \sim N \left( (\mu - \sigma^2 / 2) \mathbf{1} \, dt, \boldsymbol{\Sigma} \, dt \right),
\end{equation}
where the covariance matrix $ \boldsymbol{\Sigma}$ is given by
\begin{equation}
\Sigma_{ij} = \sigma^2 \left(\delta_{ij} + (1 - \delta_{ij}) \rho \right),
\end{equation}
with $\delta_{ij}$ being the Kronecker delta.
That is to say, in a given regime we choose the means, variances, and correlations all to be equal for the purposes of simplicity only, so that, analogously to the 1d case, a regime can be characterised in terms of the parameters
\begin{equation}
\Theta = (\mu, \sigma),
\end{equation}
in addition to a correlation $\rho$. We use the same `bullish' and `bearish' parameters as previously, $\Theta_\mathrm{bull}$ and $\Theta_\mathrm{bear}$, as well as the same regime locations and number of data points.

\begin{figure}[t!]
\centering
\includegraphics[width=\linewidth]{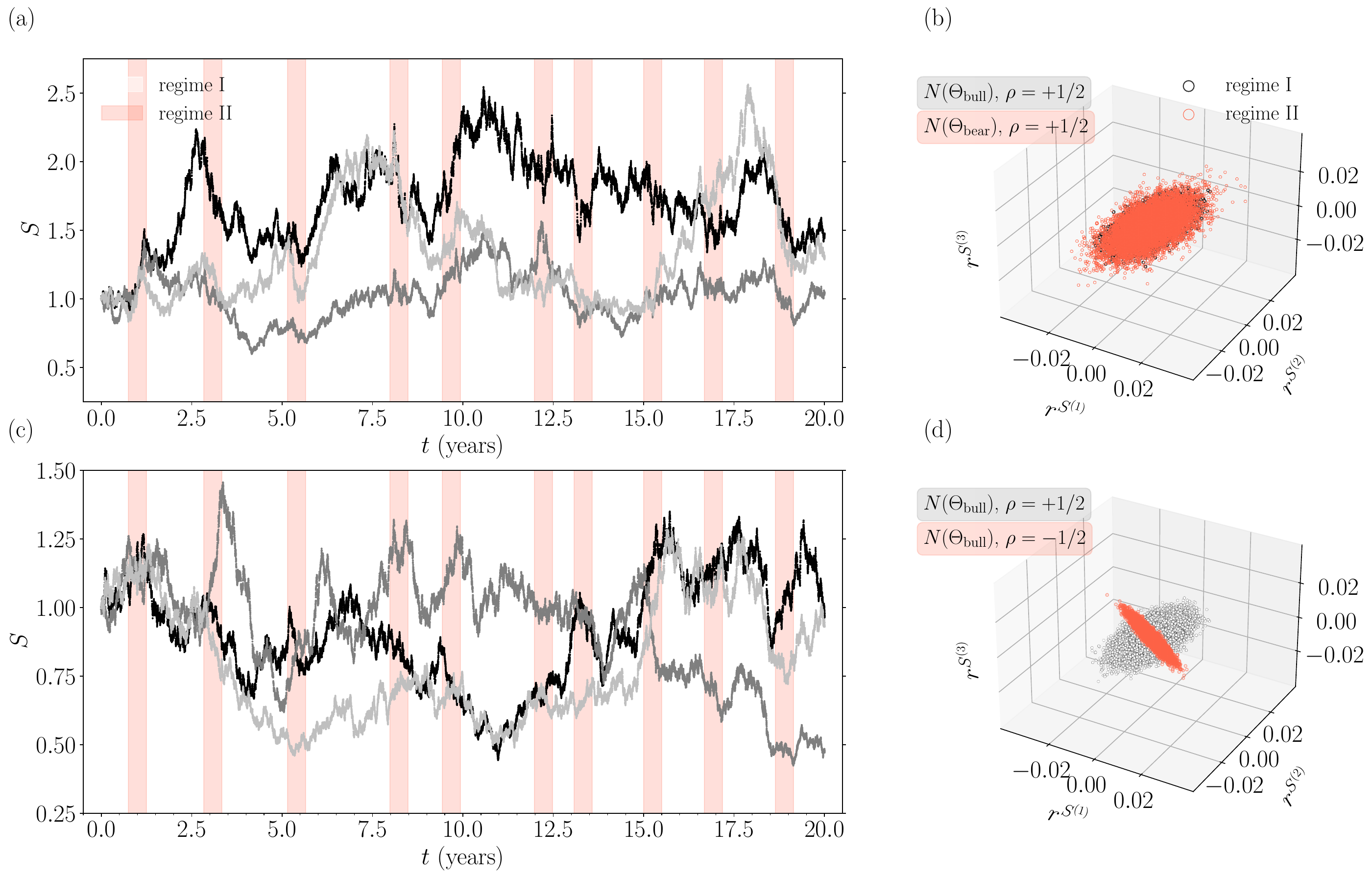}
\caption{
Synthetic 3d time series data with two regimes. (a), (c)  The time series $S(t)$, with the majority (I) and minority (II) regimes indicated. (b), (d) The empirical distributions of log returns $r^S$ corresponding to (a) and~(c) respectively. There are \mbox{$20 \times 252 \times 7 = 35,280$} data points. 
The data in (a) and~(b) has regime~I corresponding to `bullish' parameters $\Theta_\mathrm{bull}$, regime~II corresponding to `bearish' parameters $\Theta_\mathrm{bear}$, and $\rho = +1/2$ for both regimes.
The data in (c) and~(d) have regime~I and II both corresponding to `bullish' parameters $\Theta_\mathrm{bull}$, but regime~I having $\rho = +1/2$ and regime~II having $\rho = -1/2$.
The light-coloured points in the \textls[20]{distributions in (b) and~(d) correspond to the majority regime (I) periods with no highlighting in (a) and~(c); the orange points correspond to the minority regime (II) periods highlighted in orange.}
}
\label{fig:ts3-syn}
\end{figure}

\begin{figure}[t!]
\centering
\includegraphics[width=\linewidth]{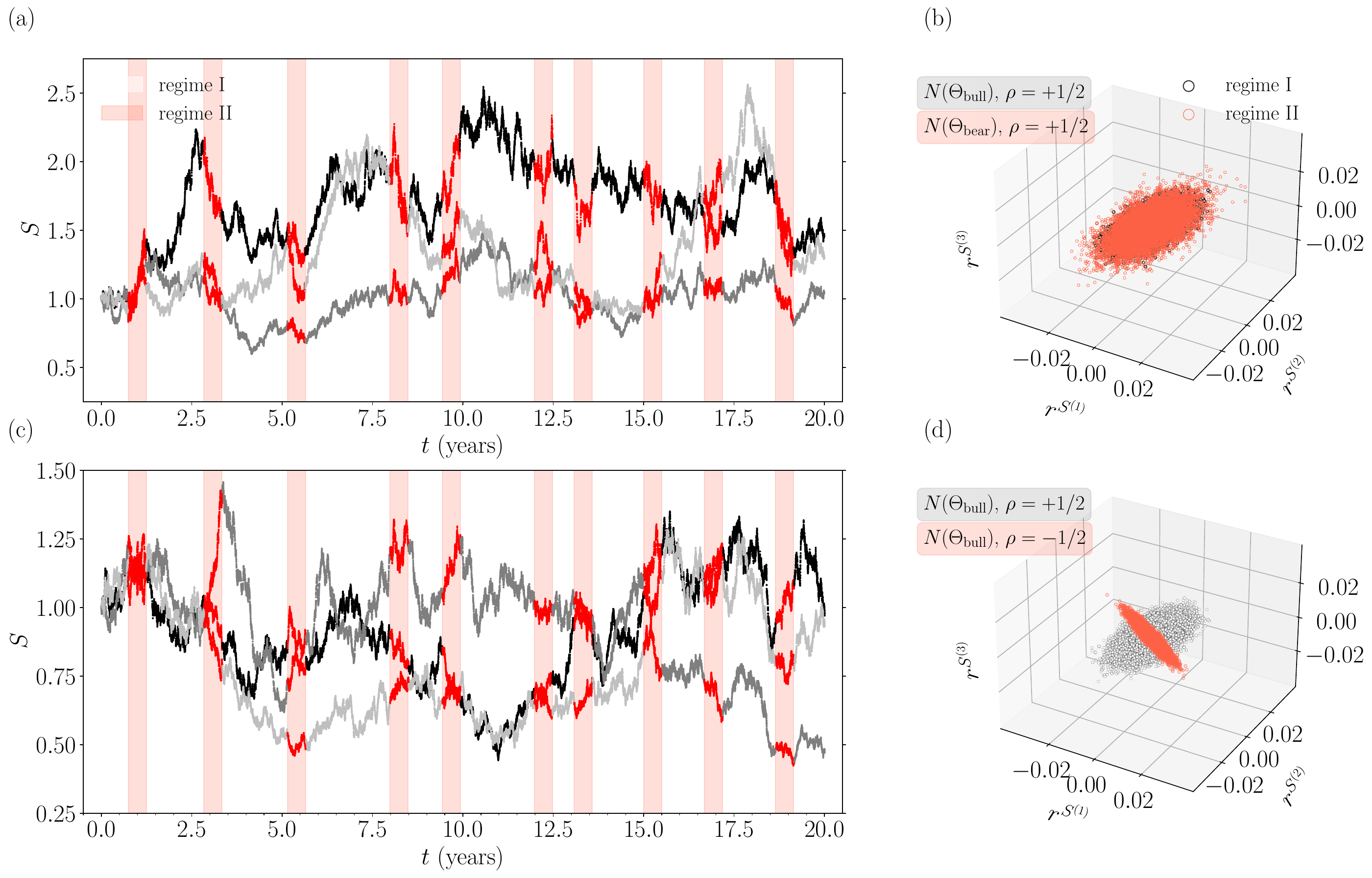}
\caption{
Results of the sWk-means algorithm for synthetic 3d time series data with two regimes shown in Figure~\ref{fig:ts3-syn}. The colouration of the points in the time series reflects the cluster assigned by the algorithm.
The window size is $h_1=60$, the lifting size is $h_2 = 12 \, (20\%)$, the number of projections is  $L=16$, and the number of clusters is $K=2$.
The run with the largest final mean centroid--centroid distance $  \langle \mathcal{W}_p(\bar{\mu}_k, \bar{\mu}_{k'}) \rangle_{k,k'} $ from 100 independent runs with different random initialisations was chosen.
}
\label{fig:ts3-syn-results}
\end{figure}

Examples of 3d synthetic data  $S(t)$ constructed in this manner can be seen in Figure~\ref{fig:ts3-syn}, along with the corresponding distributions of log returns $r^S_t$. 
The paths $S(t)$ in Figure~\ref{fig:ts3-syn}(a) contain two regimes, with the majority regime~(I) being characterised by `bullish' parameters $\Theta_\mathrm{bull}$ and the minority regimes~(II) being characterised by `bearish' parameters $\Theta_\mathrm{bear}$. Both regimes exhibit correlations \mbox{$\rho = +1/2$}.
The paths $S(t)$ in Figure~\ref{fig:ts3-syn}(c) also contain two regimes, with the majority regime~(I) and minority regime (II) both being characterised by `bullish' parameters $\Theta_\mathrm{bull}$; however, here, the majority regime (I) has correlations $\rho=+1/2$ and the minority regime (II) has correlations $\rho = -1/2$.

We now turn to discussing the results of the sWk-means clustering algorithm with these data.

\subsubsection{Results}
\label{sec:3d_ts-synthetic_results}

In this section we describe the results of the sWk-means clustering algorithm on the synthetic 3d time series data generated as just described.
As before, we chose the run with the largest final mean centroid--centroid distance $ \langle \mathcal{W}_p(\bar{\mu}_k, \bar{\mu}_{k'}) \rangle_{k,k'} $ from 100 independent runs with different random initialisations. 

Figure~\ref{fig:ts3-syn-results} shows the results of the clustering algorithm applied to the data illustrated in Figure~\ref{fig:ts3-syn}.
We use a window size of $h_1=60$ and a lifting size of $h_2 = 12 \, (20\%)$. The number of projections used is $L=9$, and the number of clusters is $K=2$.
\newpage

As is clear from the figure, the sWk-means algorithm is very effective at clustering the two regimes in the data. 
In this respect, the results for the 3d synthetic data are similar to the results for 2d synthetic data, and the algorithm continues to perform well. 
This gives us confidence that our algorithm works as expected when increasing the dimension $d$. However, with the fixed grid of projection vectors $\{ \theta^l \}$ that we use, the sWk-means algorithm  suffers from the curse of dimensionality, since in order to keep the density of points defined by the intersection of the projection vectors and the unit sphere $\mathbb{S}^{d-1}$ (and thus the accuracy of the sliced approximation to the Wasserstein distance) constant when increasing $d$, we require a number of vectors $L$ scaling with exponent ${d-1}$. This could be alleviated by randomly sampling $\theta^l$ via the Monte Carlo method, but such a choice leads to its own tradeoffs in terms of implementation, and an investigation of this falls outside the scope of this paper.

\subsection{Results on real-world financial data}
\label{sec:2d_ts-real-data}

In this section, we end by illustrating the results of the sWk-means algorithm applied to real-world financial time series data, using publicly available FX spot exchange rate data\footnotemark~as a case study. Specifically, we apply the algorithm to combined hourly Dollar-Yen (\textsc{usdjpy}) and Sterling-Dollar (\textsc{gbpusd}) spot rate data starting from 30~April~2007 until 8~August~2023. The dataset contains $100,879$ 2d data points.
\footnotetext{Specifically, we use the FX spot rate data that are available at \verb|https://www.dukascopy.com/datafeed/|.}

We choose to use $K=3$ clusters in order to give the algorithm a chance in teasing out information from the 2d dataset beyond the most obvious high- and low-variance regimes that are typically identified when using $K=2$ clusters even for 1d data. We anticipate that the additional degree of freedom will allow the algorithm to say something useful about the joint distribution of the time series in addition to the marginal behaviour.

\begin{figure}[H]
\centering
\includegraphics[width=\linewidth]{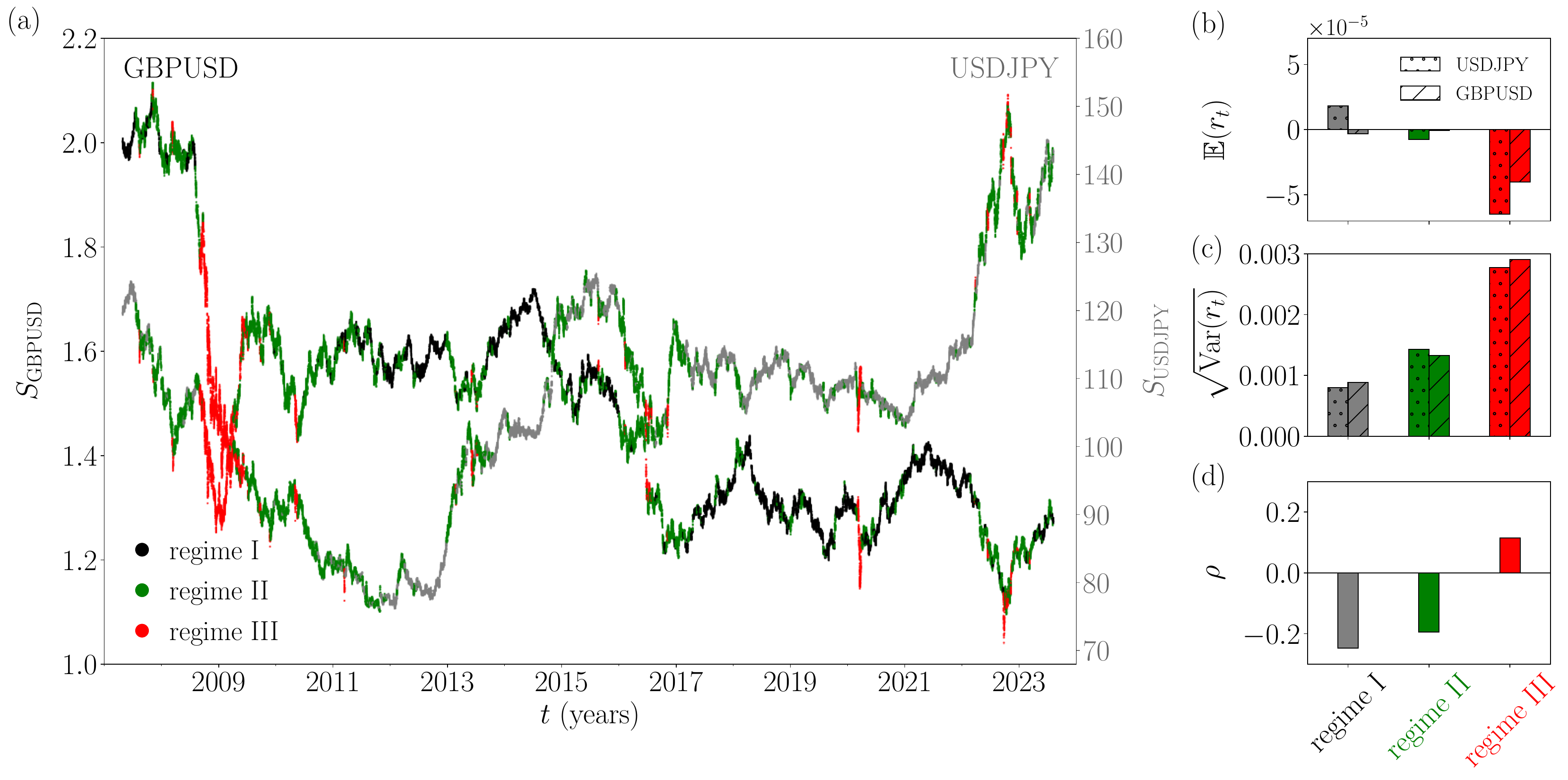}
\caption{Results of the sWk-means algorithm on 2d real-world financial time series data (combined hourly \mbox{\textsc{usdjpy}} and \mbox{\textsc{gbpusd}} spot rates from 30~April~2007 until 8~August~2023).
The window size is $h_1=60$, the lifting size is $h_2 = 12 \, (20\%)$, the number of projections is  $L=16$, and the number of clusters is $K=3$.
The run with the largest final mean centroid--centroid distance $  \langle \mathcal{W}_p(\bar{\mu}_k, \bar{\mu}_{k'}) \rangle_{k,k'} $ from 100 independent runs with different random initialisations was chosen.
(a) The dataset with points coloured according to the cluster assigned by the algorithm (I, II, or III). The algorithm is successful in identifying visibly distinct regimes in the data. (b), (c), (d) Histograms showing descriptive statistics of the returns $r_t$ in each of the regimes I, II, and III. Panel~(b) shows the average returns $\mathbb{E}(r_t)$; (c) the standard deviation of returns $\sqrt{\mathrm{Var}(r_t)}$, and~(d) the correlation~$\rho$. The coloration of the histogram bars reflects that of the corresponding cluster in Panel (a).
}
\label{fig:ts2-real-data}
\end{figure}

The dataset including the results of the sWk-means clustering algorithm can be seen in Figure~\ref{fig:ts2-real-data}(a). We use a window size $h_1=60$ and a lifting size $h_2=12 \, (20 \%)$, and, as usual, choose the clustering that maximises the mean centroid--centroid distance $  \langle \mathcal{W}_p(\bar{\mu}_k, \bar{\mu}_{k'}) \rangle_{k,k'} $ from 100 random initialisations. Each point in the time series is coloured according to the cluster (I, II, or III) assigned by the algorithm (grey/black, green, and red, respectively). 

As can be seen in the figure, the algorithm is successful in identifying visibly distinct regimes in the data. By eye, regime~III (in red) clearly corresponds to a regime exhibiting high volatility, as well as negative returns, in both \textsc{usdjpy} and \textsc{gbpusd}. Some periods corresponding to this jointly stressed regime clearly coincide with the global financial crisis (GFC) and the COVID-19 pandemic (both of which affected both currency pairs); however, it is interesting to note that periods with stresses primarily affecting only one of the currency pairs (e.g.,~the Brexit referendum or Bank of Japan machinations) have less tendency to be categorised as belonging to this jointly stressed regime, except for short periods when the stresses happen to coincide (as occurred around the Truss--Kwarteng `mini' budget, for example), and even then less clearly or only for short periods.

Regime~I (grey/black) and II (green) clearly correspond to more benign periods; by eye, it is possible to guess that regime~II (green) is more volatile than regime~I (grey/black); beyond that, their defining characteristics are less apparent. We will proceed to show that these regimes, however, exhibit meaningful differences.

In order to gain a better understanding of all three regimes identified in the absence of ground-truth labels, we calculate some descriptive statistics of the returns $r_t$ in each of the regimes $k$,{ $\{ r_{t_i} : y_{t_i} = k \}$}, and plot these in the form of the histograms shown in Figure~\ref{fig:ts2-real-data}(b), (c), and~(d). Panel~(b) shows the average returns $\mathbb{E}(r_t)$; Panel~(c) shows the standard deviation of returns $\sqrt{\mathrm{Var}(r_t)}$, and Panel~(d) shows the correlation $\rho$ between the returns $\{r_t^{S_\textsc{usdjpy}} \}$ and $\{r_t^{S_\textsc{gbpusd}} \}$ in each regime. The histogram bars are coloured according to the corresponding regime in Figure~\ref{fig:ts2-real-data}(a), and in panels~(b) and~(c), the histogram bars corresponding to \textsc{usdjpy} and \textsc{gbpusd} are indicated by dots and hatching, respectively. 

As expected, regime~III (in red) can be seen to exhibit large negative returns for both \textsc{usdjpy} and \textsc{gbpusd} (see Panel~(b)), in addition to large standard deviations (see Panel~(c)). Equally, the histogram in Panel~(c) shows that regime~II (in green) has a larger standard deviation of returns than regime~I (in grey/black). However, whilst in regime~II (green), the average returns are negative (but small) for both \textsc{usdjpy} and \textsc{gbpusd}, in regime~I, the average returns are positive for \textsc{usdjpy} and negative for \textsc{gbpusd}. 
In the more benign regimes (I and II), the returns exhibit negative correlations (see Panel~(d)); however, in the high-variance (stressed) regime (III), the correlations are instead positive (and the returns of both time series are large and negative on average). Thus, we see that when applied to these real-world \textls[-15]{financial time series data, the sWk-means algorithm is able to identify distinct regimes that exhibit obvious differences in addition to relatively subtle and diverse behaviour beyond what is easily visible by eye.}

\section{Discussion}

In this paper, we have studied, in detail, the behaviour of the Wk-means algorithm proposed in \cite{horvath2021clustering} applied to 1d time series data, and formulated an extension of the algorithm to multidimensional time series data, by approximating the multidimensional Wasserstein distance in terms of a sum of distances of 1d projection vectors -- a sliced Wasserstein distance. We call the resulting method `sliced Wasserstein k-means (sWk-means) clustering'. Using a grid of fixed projections throughout the algorithm simplifies the implementation and reduces the computational cost. 

Our particular choice of using a grid of projection vectors means that the implementation suffers from the curse of dimensionality, since in order to keep the accuracy of the sliced approximation to the full Wasserstein distance constant, a number of vectors scaling with an exponent~$d-1$ is required. Accordingly, this particular choice is expected to be suitable for multidimensional time series data where the dimension $d$ is not too large. We have shown that the algorithm performs well in two and three dimensions, with a modest number of projection vectors, and we expect the performance to extend to higher dimensions -- the fundamental method itself has no reason to deteriorate as the dimension is increased, but the computational cost will eventually become intractable. A Monte Carlo approach could be used to partially avoid the curse of dimensionality, but this comes with its own tradeoffs in terms of implementation, and we defer an investigation of this alternative to future research.

By constructing synthetic datasets, we have shown that the sWk-means algorithm performs well when applied to synthetic 2d and 3d time series data, and, in particular, can capture subtle differences between regimes whose distributions otherwise exhibit the same means and covariances. 

We ended our study by applying the sWk-means algorithm to 2d real-world financial time series data, using publicly available FX spot rate data as a case study. The algorithm is effective in identifying distinct regimes in the data whose characteristics can be analysed {\emph{a posteriori}}, including via descriptive statistics, for example. This demonstrates that our method is useful to practitioners in principle.

Another interesting extension of this technique would be to identify turning points in time series. For example, the distance to the identified cluster centroids can be used to calculate an implied probability of being in one regime or another, which changes as more information arrives and can indicate regime changes in real time; alternatively, since points in the time series belong to more than one sequence in general, instead of our majority-voting rule, a “fuzzy” measure of membership to a given class could be used and again could identify turning points (changes in regime) in real time. We propose this as an interesting avenue for future research.

In terms of alternatives to our method, contrary to the results exhibited in \cite{horvath2021clustering}, we find that hidden Markov models (HMMs) are also able to identify the regimes in some of our synthetic data, when the standardised returns are supplied to the algorithm. We infer that the unfavourable results found for HMMs applied to the synthetic data in \cite{horvath2021clustering} probably result from using something other than the standardised returns, which might be justifiable in some cases. In any case, we conclude that HMMs could be considered reasonable alternatives to the method proposed in this paper, provided that the salient details of the regimes can be captured in terms of a multivariate Gaussian, which is not always the case -- for example, for the `moon-shaped' distributions we employed.

Finally, a recent preprint by \cite{issa2023non} introduces a new nonparametric method to identify market regimes in multidimensional time series data by exploiting rough path signatures, showing good results for high dimensionality.  Signature methods for regime classification were also explored by \cite{bilokon2021market}. No doubt that nonparametric distribution- and path-based methods will continue to provide fertile ground for advances in our ability to automatically classify regimes in time series data, both in the setting of finance and beyond.

\section*{Author contributions}
 
The authors contributed equally to this paper.

\section*{Use of AI tools declaration}
The authors declare they have not used artificial intelligence (AI) tools in the creation of this article.

\section*{Acknowledgements}
We thank Daniel Mitchell for support as well as helpful discussions, especially regarding applications of the method to real-world data. 

\section*{Conflict of interest}
The authors declare no conflicts of interest in this paper.

\section*{Disclaimer}
The views expressed herein should not be considered as investment advice or promotion. They represent research undertaken by the authors and do not necessarily reflect the views of their employer, associates, or affiliates.







\end{document}

%% file: tables/ts1_accuracy_matrix_k_2.tex
\resizebox{1\textwidth}{!}{%
\begin{tabular}{lllllllllll}
\hline
 &  & \multicolumn{3}{c}{median} & \multicolumn{3}{c}{max} & \multicolumn{3}{c}{ max($\langle \mathcal{W}_p(\bar{\mu}_k, \bar{\mu}_{k'}) \rangle_{k,k'}$)} \\
 &  & 1 year & 2 years & 20 years & 1 year & 2 years & 20 years & 1 year & 2 years & 20 years \\
$h_{1}$ & $h_{2}$ &  &  &  &  &  &  &  &  &  \\
\hline
\multirow[c]{5}{*}{10} & 9 (90\%) & {\cellcolor[HTML]{FFF4B0}} \color[HTML]{000000} 55.9 & {\cellcolor[HTML]{FFF1AB}} \color[HTML]{000000} 56.6 & {\cellcolor[HTML]{FFFEC9}} \color[HTML]{000000} 52.4 & {\cellcolor[HTML]{FD863A}} \color[HTML]{F1F1F1} 76.7 & {\cellcolor[HTML]{FD8239}} \color[HTML]{F1F1F1} 77.1 & {\cellcolor[HTML]{FFFAC0}} \color[HTML]{000000} 53.7 & {\cellcolor[HTML]{FFEA9B}} \color[HTML]{000000} 58.9 & {\cellcolor[HTML]{FFF7B9}} \color[HTML]{000000} 54.7 & {\cellcolor[HTML]{FFFEC9}} \color[HTML]{000000} 52.5 \\
 & 7 (70\%) & {\cellcolor[HTML]{FFF7B7}} \color[HTML]{000000} 54.9 & {\cellcolor[HTML]{FFF0A8}} \color[HTML]{000000} 57.0 & {\cellcolor[HTML]{FFFEC9}} \color[HTML]{000000} 52.4 & {\cellcolor[HTML]{FD7234}} \color[HTML]{F1F1F1} 78.6 & {\cellcolor[HTML]{FFE691}} \color[HTML]{000000} 60.1 & {\cellcolor[HTML]{FFFBC2}} \color[HTML]{000000} 53.4 & {\cellcolor[HTML]{FFF5B5}} \color[HTML]{000000} 55.3 & {\cellcolor[HTML]{FFEDA1}} \color[HTML]{000000} 57.9 & {\cellcolor[HTML]{FFFFCC}} \color[HTML]{000000} 51.8 \\
 & 5 (50\%) & {\cellcolor[HTML]{FFF3AF}} \color[HTML]{000000} 56.0 & {\cellcolor[HTML]{FFEEA3}} \color[HTML]{000000} 57.8 & {\cellcolor[HTML]{FFFCC5}} \color[HTML]{000000} 53.0 & {\cellcolor[HTML]{F94828}} \color[HTML]{F1F1F1} 82.6 & {\cellcolor[HTML]{FFE691}} \color[HTML]{000000} 60.1 & {\cellcolor[HTML]{FFFAC0}} \color[HTML]{000000} 53.8 & {\cellcolor[HTML]{FFF8BB}} \color[HTML]{000000} 54.3 & {\cellcolor[HTML]{FFE691}} \color[HTML]{000000} 60.1 & {\cellcolor[HTML]{FFFDC6}} \color[HTML]{000000} 52.9 \\
 & 3 (30\%) & {\cellcolor[HTML]{FFF7B9}} \color[HTML]{000000} 54.7 & {\cellcolor[HTML]{FFF3AE}} \color[HTML]{000000} 56.3 & {\cellcolor[HTML]{FFFDC8}} \color[HTML]{000000} 52.6 & {\cellcolor[HTML]{FFEC9D}} \color[HTML]{000000} 58.5 & {\cellcolor[HTML]{FFE794}} \color[HTML]{000000} 59.7 & {\cellcolor[HTML]{FFFCC5}} \color[HTML]{000000} 53.1 & {\cellcolor[HTML]{FFEFA5}} \color[HTML]{000000} 57.3 & {\cellcolor[HTML]{FFF4B0}} \color[HTML]{000000} 55.8 & {\cellcolor[HTML]{FFFFCC}} \color[HTML]{000000} 51.7 \\
 & 1 (10\%) & {\cellcolor[HTML]{FFF3AE}} \color[HTML]{000000} 56.2 & {\cellcolor[HTML]{FFEFA4}} \color[HTML]{000000} 57.5 & {\cellcolor[HTML]{FFFEC9}} \color[HTML]{000000} 52.4 & {\cellcolor[HTML]{FFEEA3}} \color[HTML]{000000} 57.7 & {\cellcolor[HTML]{FFEFA4}} \color[HTML]{000000} 57.5 & {\cellcolor[HTML]{FFFEC9}} \color[HTML]{000000} 52.4 & {\cellcolor[HTML]{FFEEA3}} \color[HTML]{000000} 57.7 & {\cellcolor[HTML]{FFEFA5}} \color[HTML]{000000} 57.3 & {\cellcolor[HTML]{FFFEC9}} \color[HTML]{000000} 52.4 \\
\hline
\multirow[c]{5}{*}{20} & 18 (90\%) & {\cellcolor[HTML]{FFEA9B}} \color[HTML]{000000} 58.9 & {\cellcolor[HTML]{FFE590}} \color[HTML]{000000} 60.3 & {\cellcolor[HTML]{FFF9BE}} \color[HTML]{000000} 53.9 & {\cellcolor[HTML]{C50624}} \color[HTML]{F1F1F1} 92.5 & {\cellcolor[HTML]{D41020}} \color[HTML]{F1F1F1} 90.3 & {\cellcolor[HTML]{D41020}} \color[HTML]{F1F1F1} 90.4 & {\cellcolor[HTML]{FC6430}} \color[HTML]{F1F1F1} 79.9 & {\cellcolor[HTML]{F13624}} \color[HTML]{F1F1F1} 84.8 & {\cellcolor[HTML]{D41020}} \color[HTML]{F1F1F1} 90.4 \\
 & 14 (70\%) & {\cellcolor[HTML]{FFE998}} \color[HTML]{000000} 59.2 & {\cellcolor[HTML]{FFEC9F}} \color[HTML]{000000} 58.3 & {\cellcolor[HTML]{FFFCC5}} \color[HTML]{000000} 53.0 & {\cellcolor[HTML]{C20325}} \color[HTML]{F1F1F1} 93.2 & {\cellcolor[HTML]{D10E21}} \color[HTML]{F1F1F1} 90.8 & {\cellcolor[HTML]{D10E21}} \color[HTML]{F1F1F1} 90.7 & {\cellcolor[HTML]{F13624}} \color[HTML]{F1F1F1} 84.7 & {\cellcolor[HTML]{DF171D}} \color[HTML]{F1F1F1} 88.6 & {\cellcolor[HTML]{D10E21}} \color[HTML]{F1F1F1} 90.7 \\
 & 10 (50\%) & {\cellcolor[HTML]{FFE793}} \color[HTML]{000000} 59.9 & {\cellcolor[HTML]{FEE289}} \color[HTML]{000000} 61.3 & {\cellcolor[HTML]{FFF9BD}} \color[HTML]{000000} 54.1 & {\cellcolor[HTML]{AE0026}} \color[HTML]{F1F1F1} 95.4 & {\cellcolor[HTML]{C00225}} \color[HTML]{F1F1F1} 93.5 & {\cellcolor[HTML]{BB0026}} \color[HTML]{F1F1F1} 94.0 & {\cellcolor[HTML]{FD9841}} \color[HTML]{000000} 74.3 & {\cellcolor[HTML]{D30F20}} \color[HTML]{F1F1F1} 90.5 & {\cellcolor[HTML]{C10325}} \color[HTML]{F1F1F1} 93.4 \\
 & 6 (30\%) & {\cellcolor[HTML]{FFEFA4}} \color[HTML]{000000} 57.6 & {\cellcolor[HTML]{FEE288}} \color[HTML]{000000} 61.4 & {\cellcolor[HTML]{FFF9BD}} \color[HTML]{000000} 54.1 & {\cellcolor[HTML]{B90026}} \color[HTML]{F1F1F1} 94.2 & {\cellcolor[HTML]{C40524}} \color[HTML]{F1F1F1} 92.7 & {\cellcolor[HTML]{CB0A22}} \color[HTML]{F1F1F1} 91.7 & {\cellcolor[HTML]{CA0923}} \color[HTML]{F1F1F1} 91.9 & {\cellcolor[HTML]{CF0C21}} \color[HTML]{F1F1F1} 91.1 & {\cellcolor[HTML]{CB0A22}} \color[HTML]{F1F1F1} 91.7 \\
 & 2 (10\%) & {\cellcolor[HTML]{FFE998}} \color[HTML]{000000} 59.3 & {\cellcolor[HTML]{FFE793}} \color[HTML]{000000} 60.0 & {\cellcolor[HTML]{FFF8BA}} \color[HTML]{000000} 54.5 & {\cellcolor[HTML]{B90026}} \color[HTML]{F1F1F1} 94.2 & {\cellcolor[HTML]{CA0923}} \color[HTML]{F1F1F1} 91.8 & {\cellcolor[HTML]{BD0026}} \color[HTML]{F1F1F1} 93.9 & {\cellcolor[HTML]{B90026}} \color[HTML]{F1F1F1} 94.2 & {\cellcolor[HTML]{CA0923}} \color[HTML]{F1F1F1} 91.8 & {\cellcolor[HTML]{BD0026}} \color[HTML]{F1F1F1} 93.9 \\
\hline
\multirow[c]{5}{*}{30} & 27 (90\%) & {\cellcolor[HTML]{FEAF4B}} \color[HTML]{000000} 70.4 & {\cellcolor[HTML]{FC6430}} \color[HTML]{F1F1F1} 79.8 & {\cellcolor[HTML]{FEDF83}} \color[HTML]{000000} 62.2 & {\cellcolor[HTML]{950026}} \color[HTML]{F1F1F1} 97.9 & {\cellcolor[HTML]{A40026}} \color[HTML]{F1F1F1} 96.4 & {\cellcolor[HTML]{AC0026}} \color[HTML]{F1F1F1} 95.5 & {\cellcolor[HTML]{F33B25}} \color[HTML]{F1F1F1} 84.2 & {\cellcolor[HTML]{D6111F}} \color[HTML]{F1F1F1} 89.9 & {\cellcolor[HTML]{AC0026}} \color[HTML]{F1F1F1} 95.5 \\
 & 21 (70\%) & {\cellcolor[HTML]{FEA647}} \color[HTML]{000000} 71.9 & {\cellcolor[HTML]{FC552C}} \color[HTML]{F1F1F1} 81.4 & {\cellcolor[HTML]{FFE590}} \color[HTML]{000000} 60.4 & {\cellcolor[HTML]{970026}} \color[HTML]{F1F1F1} 97.6 & {\cellcolor[HTML]{B20026}} \color[HTML]{F1F1F1} 95.1 & {\cellcolor[HTML]{B20026}} \color[HTML]{F1F1F1} 95.1 & {\cellcolor[HTML]{FC4F2A}} \color[HTML]{F1F1F1} 81.9 & {\cellcolor[HTML]{C80723}} \color[HTML]{F1F1F1} 92.3 & {\cellcolor[HTML]{B20026}} \color[HTML]{F1F1F1} 95.1 \\
 & 15 (50\%) & {\cellcolor[HTML]{FD9841}} \color[HTML]{000000} 74.2 & {\cellcolor[HTML]{EB2B21}} \color[HTML]{F1F1F1} 86.0 & {\cellcolor[HTML]{FEDD7F}} \color[HTML]{000000} 62.7 & {\cellcolor[HTML]{950026}} \color[HTML]{F1F1F1} 97.9 & {\cellcolor[HTML]{910026}} \color[HTML]{F1F1F1} 98.2 & {\cellcolor[HTML]{970026}} \color[HTML]{F1F1F1} 97.7 & {\cellcolor[HTML]{FD933F}} \color[HTML]{000000} 75.0 & {\cellcolor[HTML]{C70723}} \color[HTML]{F1F1F1} 92.4 & {\cellcolor[HTML]{9F0026}} \color[HTML]{F1F1F1} 96.9 \\
 & 9 (30\%) & {\cellcolor[HTML]{FEA546}} \color[HTML]{000000} 72.2 & {\cellcolor[HTML]{F23924}} \color[HTML]{F1F1F1} 84.4 & {\cellcolor[HTML]{FFE997}} \color[HTML]{000000} 59.4 & {\cellcolor[HTML]{A60026}} \color[HTML]{F1F1F1} 96.2 & {\cellcolor[HTML]{A80026}} \color[HTML]{F1F1F1} 95.9 & {\cellcolor[HTML]{9F0026}} \color[HTML]{F1F1F1} 96.9 & {\cellcolor[HTML]{DD161D}} \color[HTML]{F1F1F1} 88.8 & {\cellcolor[HTML]{C00225}} \color[HTML]{F1F1F1} 93.5 & {\cellcolor[HTML]{A20026}} \color[HTML]{F1F1F1} 96.6 \\
 & 3 (10\%) & {\cellcolor[HTML]{FD903D}} \color[HTML]{000000} 75.6 & {\cellcolor[HTML]{C70723}} \color[HTML]{F1F1F1} 92.4 & {\cellcolor[HTML]{FFE998}} \color[HTML]{000000} 59.2 & {\cellcolor[HTML]{C50624}} \color[HTML]{F1F1F1} 92.6 & {\cellcolor[HTML]{A60026}} \color[HTML]{F1F1F1} 96.1 & {\cellcolor[HTML]{9B0026}} \color[HTML]{F1F1F1} 97.3 & {\cellcolor[HTML]{CB0A22}} \color[HTML]{F1F1F1} 91.7 & {\cellcolor[HTML]{C50624}} \color[HTML]{F1F1F1} 92.5 & {\cellcolor[HTML]{9B0026}} \color[HTML]{F1F1F1} 97.3 \\
\hline
\multirow[c]{6}{*}{35} & 31 (90\%) & {\cellcolor[HTML]{F74327}} \color[HTML]{F1F1F1} 83.3 & {\cellcolor[HTML]{D30F20}} \color[HTML]{F1F1F1} 90.5 & {\cellcolor[HTML]{B00026}} \color[HTML]{F1F1F1} 95.2 & {\cellcolor[HTML]{930026}} \color[HTML]{F1F1F1} 98.0 & {\cellcolor[HTML]{990026}} \color[HTML]{F1F1F1} 97.4 & {\cellcolor[HTML]{9D0026}} \color[HTML]{F1F1F1} 97.1 & {\cellcolor[HTML]{FD8F3D}} \color[HTML]{F1F1F1} 75.7 & {\cellcolor[HTML]{D41020}} \color[HTML]{F1F1F1} 90.4 & {\cellcolor[HTML]{AA0026}} \color[HTML]{F1F1F1} 95.7 \\
 & 24 (70\%) & {\cellcolor[HTML]{EE3122}} \color[HTML]{F1F1F1} 85.2 & {\cellcolor[HTML]{C40524}} \color[HTML]{F1F1F1} 92.8 & {\cellcolor[HTML]{AC0026}} \color[HTML]{F1F1F1} 95.5 & {\cellcolor[HTML]{8D0026}} \color[HTML]{F1F1F1} 98.6 & {\cellcolor[HTML]{950026}} \color[HTML]{F1F1F1} 97.9 & {\cellcolor[HTML]{A60026}} \color[HTML]{F1F1F1} 96.2 & {\cellcolor[HTML]{E0181D}} \color[HTML]{F1F1F1} 88.5 & {\cellcolor[HTML]{CE0C22}} \color[HTML]{F1F1F1} 91.2 & {\cellcolor[HTML]{AC0026}} \color[HTML]{F1F1F1} 95.6 \\
 & 17 (50\%) & {\cellcolor[HTML]{E7231E}} \color[HTML]{F1F1F1} 86.9 & {\cellcolor[HTML]{AC0026}} \color[HTML]{F1F1F1} 95.5 & {\cellcolor[HTML]{970026}} \color[HTML]{F1F1F1} 97.6 & {\cellcolor[HTML]{840026}} \color[HTML]{F1F1F1} 99.5 & {\cellcolor[HTML]{8D0026}} \color[HTML]{F1F1F1} 98.6 & {\cellcolor[HTML]{8D0026}} \color[HTML]{F1F1F1} 98.5 & {\cellcolor[HTML]{DF171D}} \color[HTML]{F1F1F1} 88.6 & {\cellcolor[HTML]{D30F20}} \color[HTML]{F1F1F1} 90.6 & {\cellcolor[HTML]{9B0026}} \color[HTML]{F1F1F1} 97.3 \\
 & 10 (30\%) & {\cellcolor[HTML]{D10E21}} \color[HTML]{F1F1F1} 90.8 & {\cellcolor[HTML]{B00026}} \color[HTML]{F1F1F1} 95.3 & {\cellcolor[HTML]{9D0026}} \color[HTML]{F1F1F1} 97.1 & {\cellcolor[HTML]{9F0026}} \color[HTML]{F1F1F1} 96.9 & {\cellcolor[HTML]{9B0026}} \color[HTML]{F1F1F1} 97.2 & {\cellcolor[HTML]{970026}} \color[HTML]{F1F1F1} 97.6 & {\cellcolor[HTML]{9F0026}} \color[HTML]{F1F1F1} 96.9 & {\cellcolor[HTML]{BB0026}} \color[HTML]{F1F1F1} 94.1 & {\cellcolor[HTML]{9D0026}} \color[HTML]{F1F1F1} 97.1 \\
 & 7 (20\%) & {\cellcolor[HTML]{C90823}} \color[HTML]{F1F1F1} 92.0 & {\cellcolor[HTML]{A60026}} \color[HTML]{F1F1F1} 96.2 & {\cellcolor[HTML]{970026}} \color[HTML]{F1F1F1} 97.7 & {\cellcolor[HTML]{AC0026}} \color[HTML]{F1F1F1} 95.6 & {\cellcolor[HTML]{930026}} \color[HTML]{F1F1F1} 98.1 & {\cellcolor[HTML]{930026}} \color[HTML]{F1F1F1} 98.0 & {\cellcolor[HTML]{B70026}} \color[HTML]{F1F1F1} 94.4 & {\cellcolor[HTML]{B60026}} \color[HTML]{F1F1F1} 94.7 & {\cellcolor[HTML]{950026}} \color[HTML]{F1F1F1} 97.9 \\
 & 3 (10\%) & {\cellcolor[HTML]{C20325}} \color[HTML]{F1F1F1} 93.2 & {\cellcolor[HTML]{A20026}} \color[HTML]{F1F1F1} 96.5 & {\cellcolor[HTML]{950026}} \color[HTML]{F1F1F1} 97.8 & {\cellcolor[HTML]{B70026}} \color[HTML]{F1F1F1} 94.4 & {\cellcolor[HTML]{970026}} \color[HTML]{F1F1F1} 97.6 & {\cellcolor[HTML]{950026}} \color[HTML]{F1F1F1} 97.9 & {\cellcolor[HTML]{B70026}} \color[HTML]{F1F1F1} 94.4 & {\cellcolor[HTML]{A80026}} \color[HTML]{F1F1F1} 95.9 & {\cellcolor[HTML]{950026}} \color[HTML]{F1F1F1} 97.8 \\
\hline
\multirow[c]{5}{*}{40} & 36 (90\%) & {\cellcolor[HTML]{D41020}} \color[HTML]{F1F1F1} 90.4 & {\cellcolor[HTML]{B90026}} \color[HTML]{F1F1F1} 94.2 & {\cellcolor[HTML]{A10026}} \color[HTML]{F1F1F1} 96.7 & {\cellcolor[HTML]{860026}} \color[HTML]{F1F1F1} 99.3 & {\cellcolor[HTML]{840026}} \color[HTML]{F1F1F1} 99.5 & {\cellcolor[HTML]{9B0026}} \color[HTML]{F1F1F1} 97.3 & {\cellcolor[HTML]{CB0A22}} \color[HTML]{F1F1F1} 91.6 & {\cellcolor[HTML]{C90823}} \color[HTML]{F1F1F1} 92.0 & {\cellcolor[HTML]{A40026}} \color[HTML]{F1F1F1} 96.3 \\
 & 28 (70\%) & {\cellcolor[HTML]{CF0C21}} \color[HTML]{F1F1F1} 91.0 & {\cellcolor[HTML]{B20026}} \color[HTML]{F1F1F1} 95.0 & {\cellcolor[HTML]{A20026}} \color[HTML]{F1F1F1} 96.6 & {\cellcolor[HTML]{880026}} \color[HTML]{F1F1F1} 99.2 & {\cellcolor[HTML]{930026}} \color[HTML]{F1F1F1} 98.0 & {\cellcolor[HTML]{9B0026}} \color[HTML]{F1F1F1} 97.3 & {\cellcolor[HTML]{C10325}} \color[HTML]{F1F1F1} 93.4 & {\cellcolor[HTML]{B70026}} \color[HTML]{F1F1F1} 94.4 & {\cellcolor[HTML]{A40026}} \color[HTML]{F1F1F1} 96.3 \\
 & 20 (50\%) & {\cellcolor[HTML]{C00225}} \color[HTML]{F1F1F1} 93.6 & {\cellcolor[HTML]{9D0026}} \color[HTML]{F1F1F1} 97.0 & {\cellcolor[HTML]{910026}} \color[HTML]{F1F1F1} 98.2 & {\cellcolor[HTML]{800026}} \color[HTML]{F1F1F1} 100.0 & {\cellcolor[HTML]{8B0026}} \color[HTML]{F1F1F1} 98.8 & {\cellcolor[HTML]{8B0026}} \color[HTML]{F1F1F1} 98.8 & {\cellcolor[HTML]{CA0923}} \color[HTML]{F1F1F1} 91.8 & {\cellcolor[HTML]{AA0026}} \color[HTML]{F1F1F1} 95.7 & {\cellcolor[HTML]{930026}} \color[HTML]{F1F1F1} 98.1 \\
 & 12 (30\%) & {\cellcolor[HTML]{B40026}} \color[HTML]{F1F1F1} 94.8 & {\cellcolor[HTML]{9F0026}} \color[HTML]{F1F1F1} 96.9 & {\cellcolor[HTML]{950026}} \color[HTML]{F1F1F1} 97.9 & {\cellcolor[HTML]{800026}} \color[HTML]{F1F1F1} 99.9 & {\cellcolor[HTML]{860026}} \color[HTML]{F1F1F1} 99.4 & {\cellcolor[HTML]{8F0026}} \color[HTML]{F1F1F1} 98.4 & {\cellcolor[HTML]{B00026}} \color[HTML]{F1F1F1} 95.2 & {\cellcolor[HTML]{AC0026}} \color[HTML]{F1F1F1} 95.5 & {\cellcolor[HTML]{950026}} \color[HTML]{F1F1F1} 97.8 \\
 & 4 (10\%) & {\cellcolor[HTML]{AE0026}} \color[HTML]{F1F1F1} 95.4 & {\cellcolor[HTML]{950026}} \color[HTML]{F1F1F1} 97.8 & {\cellcolor[HTML]{8D0026}} \color[HTML]{F1F1F1} 98.6 & {\cellcolor[HTML]{9B0026}} \color[HTML]{F1F1F1} 97.2 & {\cellcolor[HTML]{8A0026}} \color[HTML]{F1F1F1} 99.0 & {\cellcolor[HTML]{8D0026}} \color[HTML]{F1F1F1} 98.6 & {\cellcolor[HTML]{BB0026}} \color[HTML]{F1F1F1} 94.0 & {\cellcolor[HTML]{9D0026}} \color[HTML]{F1F1F1} 97.1 & {\cellcolor[HTML]{8F0026}} \color[HTML]{F1F1F1} 98.4 \\
\cline{1-11}
\hline
\end{tabular}
}

%% file: tables/ts2_k2_new_acc.tex
\resizebox{0.97\textwidth}{!}{%

\begin{tabular}{lllllllll}

\hline
 &  &  & \multicolumn{2}{c}{median} & \multicolumn{2}{c}{max} & \multicolumn{2}{c}{max($\langle \mathcal{W}_p(\bar{\mu}_k, \bar{\mu}_{k'}) \rangle_{k,k'}$)} \\
 &  &  & Type A & Type B & Type A & Type B & Type A & Type B \\
$h_1$ & $h_2$ & $ L$ &  &  &  &  &  &  \\
\hline
\multirow[c]{4}{*}{10} & \multirow[c]{4}{*}{2 (20\%)} & 2 & {\cellcolor[HTML]{FFFEC9}} \color[HTML]{000000} 50.5 & {\cellcolor[HTML]{FFFEC9}} \color[HTML]{000000} 50.5 & {\cellcolor[HTML]{FFFDC8}} \color[HTML]{000000} 50.6 & {\cellcolor[HTML]{FFFDC8}} \color[HTML]{000000} 50.7 & {\cellcolor[HTML]{FFFEC9}} \color[HTML]{000000} 50.4 & {\cellcolor[HTML]{FFFEC9}} \color[HTML]{000000} 50.4 \\
 &  & 4 & {\cellcolor[HTML]{FFFBC2}} \color[HTML]{000000} 51.4 & {\cellcolor[HTML]{FFFDC8}} \color[HTML]{000000} 50.6 & {\cellcolor[HTML]{FFFAC1}} \color[HTML]{000000} 51.7 & {\cellcolor[HTML]{FFFDC8}} \color[HTML]{000000} 50.7 & {\cellcolor[HTML]{FFFCC5}} \color[HTML]{000000} 51.1 & {\cellcolor[HTML]{FFFDC8}} \color[HTML]{000000} 50.6 \\
 &  & 9 & {\cellcolor[HTML]{FFFCC5}} \color[HTML]{000000} 51.0 & {\cellcolor[HTML]{FFFDC8}} \color[HTML]{000000} 50.6 & {\cellcolor[HTML]{FFFBC2}} \color[HTML]{000000} 51.4 & {\cellcolor[HTML]{FFFDC6}} \color[HTML]{000000} 50.8 & {\cellcolor[HTML]{FFFBC2}} \color[HTML]{000000} 51.4 & {\cellcolor[HTML]{FFFDC8}} \color[HTML]{000000} 50.7 \\
 &  & 16 & {\cellcolor[HTML]{FFFCC5}} \color[HTML]{000000} 51.1 & {\cellcolor[HTML]{FFFDC8}} \color[HTML]{000000} 50.6 & {\cellcolor[HTML]{FFFBC2}} \color[HTML]{000000} 51.5 & {\cellcolor[HTML]{FFFDC6}} \color[HTML]{000000} 50.8 & {\cellcolor[HTML]{FFFDC6}} \color[HTML]{000000} 50.9 & {\cellcolor[HTML]{FFFDC8}} \color[HTML]{000000} 50.7 \\
\hline
\multirow[c]{4}{*}{20} & \multirow[c]{4}{*}{4 (20\%)} & 2 & {\cellcolor[HTML]{FFF8BB}} \color[HTML]{000000} 52.4 & {\cellcolor[HTML]{FFFECB}} \color[HTML]{000000} 50.2 & {\cellcolor[HTML]{970026}} \color[HTML]{F1F1F1} 97.0 & {\cellcolor[HTML]{FFFDC6}} \color[HTML]{000000} 50.9 & {\cellcolor[HTML]{970026}} \color[HTML]{F1F1F1} 97.0 & {\cellcolor[HTML]{FFFECB}} \color[HTML]{000000} 50.2 \\
 &  & 4 & {\cellcolor[HTML]{FFEDA1}} \color[HTML]{000000} 56.0 & {\cellcolor[HTML]{FFF3AE}} \color[HTML]{000000} 54.3 & {\cellcolor[HTML]{930026}} \color[HTML]{F1F1F1} 97.4 & {\cellcolor[HTML]{840026}} \color[HTML]{F1F1F1} 99.0 & {\cellcolor[HTML]{950026}} \color[HTML]{F1F1F1} 97.2 & {\cellcolor[HTML]{FFEC9D}} \color[HTML]{000000} 56.7 \\
 &  & 9 & {\cellcolor[HTML]{FFF1A9}} \color[HTML]{000000} 55.0 & {\cellcolor[HTML]{FFF0A7}} \color[HTML]{000000} 55.3 & {\cellcolor[HTML]{930026}} \color[HTML]{F1F1F1} 97.4 & {\cellcolor[HTML]{840026}} \color[HTML]{F1F1F1} 99.1 & {\cellcolor[HTML]{950026}} \color[HTML]{F1F1F1} 97.2 & {\cellcolor[HTML]{FFF1A9}} \color[HTML]{000000} 55.0 \\
 &  & 16 & {\cellcolor[HTML]{FFEFA4}} \color[HTML]{000000} 55.8 & {\cellcolor[HTML]{FFEFA5}} \color[HTML]{000000} 55.6 & {\cellcolor[HTML]{930026}} \color[HTML]{F1F1F1} 97.4 & {\cellcolor[HTML]{840026}} \color[HTML]{F1F1F1} 99.1 & {\cellcolor[HTML]{950026}} \color[HTML]{F1F1F1} 97.2 & {\cellcolor[HTML]{FFF1AB}} \color[HTML]{000000} 54.8 \\
\hline
\multirow[c]{4}{*}{30} & \multirow[c]{4}{*}{6 (20\%)} & 2 & {\cellcolor[HTML]{8A0026}} \color[HTML]{F1F1F1} 98.5 & {\cellcolor[HTML]{FFFECB}} \color[HTML]{000000} 50.2 & {\cellcolor[HTML]{860026}} \color[HTML]{F1F1F1} 98.8 & {\cellcolor[HTML]{FFFAC0}} \color[HTML]{000000} 51.9 & {\cellcolor[HTML]{8A0026}} \color[HTML]{F1F1F1} 98.4 & {\cellcolor[HTML]{FFFFCC}} \color[HTML]{000000} 50.0 \\
 &  & 4 & {\cellcolor[HTML]{860026}} \color[HTML]{F1F1F1} 98.9 & {\cellcolor[HTML]{FEAD4A}} \color[HTML]{000000} 69.5 & {\cellcolor[HTML]{840026}} \color[HTML]{F1F1F1} 99.0 & {\cellcolor[HTML]{800026}} \color[HTML]{F1F1F1} 99.5 & {\cellcolor[HTML]{860026}} \color[HTML]{F1F1F1} 98.8 & {\cellcolor[HTML]{800026}} \color[HTML]{F1F1F1} 99.5 \\
 &  & 9 & {\cellcolor[HTML]{860026}} \color[HTML]{F1F1F1} 98.8 & {\cellcolor[HTML]{FD9D43}} \color[HTML]{000000} 72.1 & {\cellcolor[HTML]{840026}} \color[HTML]{F1F1F1} 99.0 & {\cellcolor[HTML]{800026}} \color[HTML]{F1F1F1} 99.5 & {\cellcolor[HTML]{860026}} \color[HTML]{F1F1F1} 98.8 & {\cellcolor[HTML]{800026}} \color[HTML]{F1F1F1} 99.5 \\
 &  & 16 & {\cellcolor[HTML]{860026}} \color[HTML]{F1F1F1} 98.8 & {\cellcolor[HTML]{FD9A42}} \color[HTML]{000000} 72.5 & {\cellcolor[HTML]{860026}} \color[HTML]{F1F1F1} 98.9 & {\cellcolor[HTML]{800026}} \color[HTML]{F1F1F1} 99.5 & {\cellcolor[HTML]{860026}} \color[HTML]{F1F1F1} 98.8 & {\cellcolor[HTML]{800026}} \color[HTML]{F1F1F1} 99.4 \\
\hline
\multirow[c]{4}{*}{35} & \multirow[c]{4}{*}{7 (20\%)} & 2 & {\cellcolor[HTML]{840026}} \color[HTML]{F1F1F1} 99.1 & {\cellcolor[HTML]{FFFBC2}} \color[HTML]{000000} 51.4 & {\cellcolor[HTML]{820026}} \color[HTML]{F1F1F1} 99.3 & {\cellcolor[HTML]{FFF9BE}} \color[HTML]{000000} 52.1 & {\cellcolor[HTML]{820026}} \color[HTML]{F1F1F1} 99.2 & {\cellcolor[HTML]{FFFFCC}} \color[HTML]{000000} 50.0 \\
 &  & 4 & {\cellcolor[HTML]{840026}} \color[HTML]{F1F1F1} 99.1 & {\cellcolor[HTML]{800026}} \color[HTML]{F1F1F1} 99.4 & {\cellcolor[HTML]{840026}} \color[HTML]{F1F1F1} 99.1 & {\cellcolor[HTML]{800026}} \color[HTML]{F1F1F1} 99.5 & {\cellcolor[HTML]{860026}} \color[HTML]{F1F1F1} 98.9 & {\cellcolor[HTML]{800026}} \color[HTML]{F1F1F1} 99.4 \\
 &  & 9 & {\cellcolor[HTML]{840026}} \color[HTML]{F1F1F1} 99.1 & {\cellcolor[HTML]{800026}} \color[HTML]{F1F1F1} 99.5 & {\cellcolor[HTML]{820026}} \color[HTML]{F1F1F1} 99.2 & {\cellcolor[HTML]{800026}} \color[HTML]{F1F1F1} 99.5 & {\cellcolor[HTML]{840026}} \color[HTML]{F1F1F1} 99.0 & {\cellcolor[HTML]{800026}} \color[HTML]{F1F1F1} 99.4 \\
 &  & 16 & {\cellcolor[HTML]{840026}} \color[HTML]{F1F1F1} 99.1 & {\cellcolor[HTML]{800026}} \color[HTML]{F1F1F1} 99.5 & {\cellcolor[HTML]{820026}} \color[HTML]{F1F1F1} 99.2 & {\cellcolor[HTML]{800026}} \color[HTML]{F1F1F1} 99.5 & {\cellcolor[HTML]{840026}} \color[HTML]{F1F1F1} 99.0 & {\cellcolor[HTML]{800026}} \color[HTML]{F1F1F1} 99.4 \\
\cline{1-9}
\hline
\end{tabular}
}

%% file: tables/ts2_k3_new_acc.tex
\resizebox{1.0\textwidth}{!}{%
\begin{tabular}{lllllllll}
\hline
 &  &  & \multicolumn{2}{c}{median} & \multicolumn{2}{c}{max} & \multicolumn{2}{c}{max($\langle \mathcal{W}_p(\bar{\mu}_k, \bar{\mu}_{k'}) \rangle_{k,k'}$)} \\
 &  &  & Type C & Type D & Type C & Type D & Type C & Type D \\
$h_1$ & $h_2$ & $L$ &  &  &  &  &  &  \\
\hline
\multirow[c]{4}{*}{20} & \multirow[c]{4}{*}{4 (20\%)} & 2 & {\cellcolor[HTML]{FFF8BB}} \color[HTML]{000000} 50.2 & {\cellcolor[HTML]{FFEFA4}} \color[HTML]{000000} 53.5 & {\cellcolor[HTML]{FFF3AE}} \color[HTML]{000000} 52.1 & {\cellcolor[HTML]{FFEC9D}} \color[HTML]{000000} 54.6 & {\cellcolor[HTML]{FFF4B2}} \color[HTML]{000000} 51.5 & {\cellcolor[HTML]{FFF3AE}} \color[HTML]{000000} 52.2 \\
 &  & 4 & {\cellcolor[HTML]{FFF5B3}} \color[HTML]{000000} 51.3 & {\cellcolor[HTML]{FFEEA3}} \color[HTML]{000000} 53.8 & {\cellcolor[HTML]{FFF2AC}} \color[HTML]{000000} 52.3 & {\cellcolor[HTML]{FFEA99}} \color[HTML]{000000} 55.3 & {\cellcolor[HTML]{FFF4B0}} \color[HTML]{000000} 51.7 & {\cellcolor[HTML]{FFEDA1}} \color[HTML]{000000} 53.9 \\
 &  & 9 & {\cellcolor[HTML]{FFF4B2}} \color[HTML]{000000} 51.6 & {\cellcolor[HTML]{FFF0A8}} \color[HTML]{000000} 53.0 & {\cellcolor[HTML]{FFF1AB}} \color[HTML]{000000} 52.5 & {\cellcolor[HTML]{FFEB9C}} \color[HTML]{000000} 54.8 & {\cellcolor[HTML]{FFF4B2}} \color[HTML]{000000} 51.6 & {\cellcolor[HTML]{FFEB9C}} \color[HTML]{000000} 54.8 \\
 &  & 16 & {\cellcolor[HTML]{FFF4B2}} \color[HTML]{000000} 51.5 & {\cellcolor[HTML]{FFF0A7}} \color[HTML]{000000} 53.1 & {\cellcolor[HTML]{FFF2AC}} \color[HTML]{000000} 52.3 & {\cellcolor[HTML]{FFEB9C}} \color[HTML]{000000} 54.9 & {\cellcolor[HTML]{FFF4B2}} \color[HTML]{000000} 51.5 & {\cellcolor[HTML]{FFEB9C}} \color[HTML]{000000} 54.9 \\
\hline
\multirow[c]{4}{*}{30} & \multirow[c]{4}{*}{6 (20\%)} & 2 & {\cellcolor[HTML]{FFF8BB}} \color[HTML]{000000} 50.1 & {\cellcolor[HTML]{FFEDA0}} \color[HTML]{000000} 54.2 & {\cellcolor[HTML]{FFF2AC}} \color[HTML]{000000} 52.3 & {\cellcolor[HTML]{FFE998}} \color[HTML]{000000} 55.4 & {\cellcolor[HTML]{FFF4B0}} \color[HTML]{000000} 51.8 & {\cellcolor[HTML]{FFEA9B}} \color[HTML]{000000} 55.0 \\
 &  & 4 & {\cellcolor[HTML]{FFF7B7}} \color[HTML]{000000} 50.8 & {\cellcolor[HTML]{FFEFA4}} \color[HTML]{000000} 53.6 & {\cellcolor[HTML]{860026}} \color[HTML]{F1F1F1} 98.9 & {\cellcolor[HTML]{D7121F}} \color[HTML]{F1F1F1} 88.6 & {\cellcolor[HTML]{860026}} \color[HTML]{F1F1F1} 98.8 & {\cellcolor[HTML]{D7121F}} \color[HTML]{F1F1F1} 88.6 \\
 &  & 9 & {\cellcolor[HTML]{FFF4B2}} \color[HTML]{000000} 51.5 & {\cellcolor[HTML]{FFF0A7}} \color[HTML]{000000} 53.1 & {\cellcolor[HTML]{860026}} \color[HTML]{F1F1F1} 98.9 & {\cellcolor[HTML]{D6111F}} \color[HTML]{F1F1F1} 88.7 & {\cellcolor[HTML]{880026}} \color[HTML]{F1F1F1} 98.7 & {\cellcolor[HTML]{D9131F}} \color[HTML]{F1F1F1} 88.4 \\
 &  & 16 & {\cellcolor[HTML]{FFF4B2}} \color[HTML]{000000} 51.5 & {\cellcolor[HTML]{FFEEA3}} \color[HTML]{000000} 53.7 & {\cellcolor[HTML]{860026}} \color[HTML]{F1F1F1} 98.9 & {\cellcolor[HTML]{D7121F}} \color[HTML]{F1F1F1} 88.5 & {\cellcolor[HTML]{880026}} \color[HTML]{F1F1F1} 98.7 & {\cellcolor[HTML]{D7121F}} \color[HTML]{F1F1F1} 88.5 \\
\hline
\multirow[c]{4}{*}{40} & \multirow[c]{4}{*}{8 (20\%)} & 2 & {\cellcolor[HTML]{FFF2AC}} \color[HTML]{000000} 52.4 & {\cellcolor[HTML]{FFEFA4}} \color[HTML]{000000} 53.5 & {\cellcolor[HTML]{E1191D}} \color[HTML]{F1F1F1} 87.0 & {\cellcolor[HTML]{CE0C22}} \color[HTML]{F1F1F1} 90.1 & {\cellcolor[HTML]{E1191D}} \color[HTML]{F1F1F1} 87.0 & {\cellcolor[HTML]{D30F20}} \color[HTML]{F1F1F1} 89.3 \\
 &  & 4 & {\cellcolor[HTML]{FFF9BE}} \color[HTML]{000000} 49.7 & {\cellcolor[HTML]{FFEDA1}} \color[HTML]{000000} 53.9 & {\cellcolor[HTML]{820026}} \color[HTML]{F1F1F1} 99.3 & {\cellcolor[HTML]{C90823}} \color[HTML]{F1F1F1} 91.0 & {\cellcolor[HTML]{820026}} \color[HTML]{F1F1F1} 99.3 & {\cellcolor[HTML]{CE0C22}} \color[HTML]{F1F1F1} 90.2 \\
 &  & 9 & {\cellcolor[HTML]{FFF9BE}} \color[HTML]{000000} 49.8 & {\cellcolor[HTML]{FFEEA3}} \color[HTML]{000000} 53.7 & {\cellcolor[HTML]{820026}} \color[HTML]{F1F1F1} 99.3 & {\cellcolor[HTML]{CA0923}} \color[HTML]{F1F1F1} 90.8 & {\cellcolor[HTML]{840026}} \color[HTML]{F1F1F1} 99.1 & {\cellcolor[HTML]{CE0C22}} \color[HTML]{F1F1F1} 90.1 \\
 &  & 16 & {\cellcolor[HTML]{FFF9BD}} \color[HTML]{000000} 50.0 & {\cellcolor[HTML]{FFEEA3}} \color[HTML]{000000} 53.8 & {\cellcolor[HTML]{820026}} \color[HTML]{F1F1F1} 99.3 & {\cellcolor[HTML]{C90823}} \color[HTML]{F1F1F1} 91.0 & {\cellcolor[HTML]{840026}} \color[HTML]{F1F1F1} 99.1 & {\cellcolor[HTML]{CE0C22}} \color[HTML]{F1F1F1} 90.2 \\
\hline
\multirow[c]{4}{*}{50} & \multirow[c]{4}{*}{10 (20\%)} & 2 & {\cellcolor[HTML]{FFF3AE}} \color[HTML]{000000} 52.2 & {\cellcolor[HTML]{FFF1AB}} \color[HTML]{000000} 52.6 & {\cellcolor[HTML]{DD161D}} \color[HTML]{F1F1F1} 87.6 & {\cellcolor[HTML]{C30424}} \color[HTML]{F1F1F1} 91.9 & {\cellcolor[HTML]{DD161D}} \color[HTML]{F1F1F1} 87.6 & {\cellcolor[HTML]{CD0B22}} \color[HTML]{F1F1F1} 90.4 \\
 &  & 4 & {\cellcolor[HTML]{FFFAC0}} \color[HTML]{000000} 49.5 & {\cellcolor[HTML]{FFEFA5}} \color[HTML]{000000} 53.3 & {\cellcolor[HTML]{800026}} \color[HTML]{F1F1F1} 99.5 & {\cellcolor[HTML]{820026}} \color[HTML]{F1F1F1} 99.2 & {\cellcolor[HTML]{800026}} \color[HTML]{F1F1F1} 99.4 & {\cellcolor[HTML]{C10325}} \color[HTML]{F1F1F1} 92.3 \\
 &  & 9 & {\cellcolor[HTML]{FFFDC6}} \color[HTML]{000000} 48.6 & {\cellcolor[HTML]{FFF0A8}} \color[HTML]{000000} 53.0 & {\cellcolor[HTML]{800026}} \color[HTML]{F1F1F1} 99.5 & {\cellcolor[HTML]{860026}} \color[HTML]{F1F1F1} 98.8 & {\cellcolor[HTML]{800026}} \color[HTML]{F1F1F1} 99.4 & {\cellcolor[HTML]{C10325}} \color[HTML]{F1F1F1} 92.4 \\
 &  & 16 & {\cellcolor[HTML]{FFFDC6}} \color[HTML]{000000} 48.5 & {\cellcolor[HTML]{FFF0A7}} \color[HTML]{000000} 53.1 & {\cellcolor[HTML]{800026}} \color[HTML]{F1F1F1} 99.5 & {\cellcolor[HTML]{840026}} \color[HTML]{F1F1F1} 99.0 & {\cellcolor[HTML]{800026}} \color[HTML]{F1F1F1} 99.4 & {\cellcolor[HTML]{BE0126}} \color[HTML]{F1F1F1} 92.8 \\
\hline
\multirow[c]{4}{*}{60} & \multirow[c]{4}{*}{12 (20\%)} & 2 & {\cellcolor[HTML]{FFEEA3}} \color[HTML]{000000} 53.8 & {\cellcolor[HTML]{FFF2AC}} \color[HTML]{000000} 52.4 & {\cellcolor[HTML]{DA141E}} \color[HTML]{F1F1F1} 88.2 & {\cellcolor[HTML]{AC0026}} \color[HTML]{F1F1F1} 94.9 & {\cellcolor[HTML]{DF171D}} \color[HTML]{F1F1F1} 87.4 & {\cellcolor[HTML]{C50624}} \color[HTML]{F1F1F1} 91.6 \\
 &  & 4 & {\cellcolor[HTML]{FFFFCC}} \color[HTML]{000000} 47.6 & {\cellcolor[HTML]{FFF2AC}} \color[HTML]{000000} 52.4 & {\cellcolor[HTML]{800026}} \color[HTML]{F1F1F1} 99.5 & {\cellcolor[HTML]{800026}} \color[HTML]{F1F1F1} 99.5 & {\cellcolor[HTML]{800026}} \color[HTML]{F1F1F1} 99.5 & {\cellcolor[HTML]{800026}} \color[HTML]{F1F1F1} 99.4 \\
 &  & 9 & {\cellcolor[HTML]{FFEFA4}} \color[HTML]{000000} 53.5 & {\cellcolor[HTML]{FFF1A9}} \color[HTML]{000000} 52.7 & {\cellcolor[HTML]{800026}} \color[HTML]{F1F1F1} 99.6 & {\cellcolor[HTML]{800026}} \color[HTML]{F1F1F1} 99.4 & {\cellcolor[HTML]{800026}} \color[HTML]{F1F1F1} 99.6 & {\cellcolor[HTML]{800026}} \color[HTML]{F1F1F1} 99.4 \\
 &  & 16 & {\cellcolor[HTML]{FFEFA5}} \color[HTML]{000000} 53.4 & {\cellcolor[HTML]{FFF1A9}} \color[HTML]{000000} 52.7 & {\cellcolor[HTML]{800026}} \color[HTML]{F1F1F1} 99.6 & {\cellcolor[HTML]{820026}} \color[HTML]{F1F1F1} 99.3 & {\cellcolor[HTML]{800026}} \color[HTML]{F1F1F1} 99.5 & {\cellcolor[HTML]{820026}} \color[HTML]{F1F1F1} 99.3 \\
\cline{1-9}
\hline
\end{tabular}
}